\documentclass[
 reprint,
superscriptaddress,
 amsmath,amssymb,
 aps,
pra,
onecolumn,
longbibliography
]{revtex4-2}

\usepackage{graphicx,wrapfig,float}
\usepackage{dcolumn}
\usepackage{bm}
\graphicspath{{preliminary_figures}}
\usepackage{color}
\usepackage{MnSymbol}
\usepackage{bbding}
\usepackage{wasysym}
\usepackage{tikz}
\usetikzlibrary{shapes}

\usepackage{comment}

\definecolor{cream}{RGB}{222,217,201}

\begin{document}

\definecolor{Blue1}{rgb}{0.38, 0.5, 0.91}
\definecolor{Blue2}{rgb}{0.55, 0.69, 0.996}
\definecolor{Blue3}{rgb}{0.72, 0.81, 0.98}
\definecolor{White}{rgb}{0.82, 0.82, 0.82}
\definecolor{Red3}{rgb}{0.957, 0.77, 0.68}
\definecolor{Red2}{rgb}{0.957, 0.6, 0.482}
\definecolor{Red1}{rgb}{0.87, 0.38, 0.3}
\definecolor{Red2d}{rgb}{0.84, 0.19, 0.15}
\definecolor{White2d}{rgb}{0.95, 0.95, 0.95}

\definecolor{Blue2d}{rgb}{0.19, 0.21, 0.58}
\definecolor{Red2d}{rgb}{0.84, 0.19, 0.15}
\definecolor{White2d}{rgb}{0.95, 0.95, 0.95}

\definecolor{vis1}{rgb}{0.32, 0.43, 0.86}
\definecolor{vis2}{rgb}{0.42, 0.55, 0.94}
\definecolor{vis3}{rgb}{0.53, 0.66, 0.99}
\definecolor{vis4}{rgb}{0.63, 0.75, 1.0}
\definecolor{vis5}{rgb}{0.73, 0.82, 0.97}
\definecolor{vis6}{rgb}{0.83, 0.86, 0.91}
\definecolor{vis7}{rgb}{0.9, 0.84, 0.81}
\definecolor{vis8}{rgb}{0.95, 0.78, 0.69}
\definecolor{vis9}{rgb}{0.97, 0.69, 0.57}
\definecolor{vis10}{rgb}{0.95, 0.57, 0.45}
\definecolor{vis11}{rgb}{0.89, 0.43, 0.34}
\definecolor{vis12}{rgb}{0.81, 0.27, 0.24}
\definecolor{Pink}{rgb}{0.92, 0.153, 0.761}

\newcommand{\visline}{\raisebox{2pt}{\tikz{\draw[-,vis1,solid,line width = 2.0pt](0,0) -- (5mm,0);}}}
\newcommand{\vislinee}{\raisebox{2pt}{\tikz{\draw[-,vis2,solid,line width = 2.0pt](0,0) -- (5mm,0);}}}
\newcommand{\vislineee}{\raisebox{2pt}{\tikz{\draw[-,vis3,solid,line width = 2.0pt](0,0) -- (5mm,0);}}}
\newcommand{\vislineeee}{\raisebox{2pt}{\tikz{\draw[-,vis4,solid,line width = 2.0pt](0,0) -- (5mm,0);}}}
\newcommand{\vislineeeee}{\raisebox{2pt}{\tikz{\draw[-,vis5,solid,line width = 2.0pt](0,0) -- (5mm,0);}}}
\newcommand{\vislineeeeee}{\raisebox{2pt}{\tikz{\draw[-,vis6,solid,line width = 2.0pt](0,0) -- (5mm,0);}}}
\newcommand{\vislineeeeeee}{\raisebox{2pt}{\tikz{\draw[-,vis7,solid,line width = 2.0pt](0,0) -- (5mm,0);}}}
\newcommand{\vislineeeeeeee}{\raisebox{2pt}{\tikz{\draw[-,vis8,solid,line width = 2.0pt](0,0) -- (5mm,0);}}}
\newcommand{\vislineeeeeeeee}{\raisebox{2pt}{\tikz{\draw[-,vis9,solid,line width = 2.0pt](0,0) -- (5mm,0);}}}
\newcommand{\vislineeeeeeeeee}{\raisebox{2pt}{\tikz{\draw[-,vis10,solid,line width = 2.0pt](0,0) -- (5mm,0);}}}
\newcommand{\vislineeeeeeeeeee}{\raisebox{2pt}{\tikz{\draw[-,vis11,solid,line width = 2.0pt](0,0) -- (5mm,0);}}}
\newcommand{\vislineeeeeeeeeeee}{\raisebox{2pt}{\tikz{\draw[-,vis12,solid,line width = 2.0pt](0,0) -- (5mm,0);}}}

\newcommand{\tikzmycircle}[2][red,fill=red]{\tikz[baseline=0ex]{\draw[#1,radius=#2] (0,0.08) circle}}
\newcommand{\stressline}{\raisebox{2pt}{\tikz{\draw[-,Blue1,solid,line width = 2.0pt](0,0) -- (5mm,0);}}}
\newcommand{\stresslinee}{\raisebox{2pt}{\tikz{\draw[-,Red1,solid,line width = 2.0pt](0,0) -- (5mm,0);}}}


\newcommand{\tikzcircle}[2][red,fill=red]{\raisebox{0pt}{\tikz[baseline=0ex]\draw[-,black,dotted,line width = 1.0pt](-0.5,0.8mm) -- (5.5mm,0.8mm);\draw[#1,radius=#2] (0,0.08) circle}}%
\newcommand{\mydiamond}{\raisebox{0pt}{\tikz{\node[draw=black, scale=0.5, diamond, fill=black!10!Red1](){};}}}

\newcommand{\blackline}{\raisebox{2pt}{\tikz{\draw[-,black!40!black,solid,line width = 1.0pt](0,0) -- (5mm,0);}}}

\title{Pressure sensitivity in non-local flow behaviour of dense hydrogel particle suspensions}

\author{Zohreh Farmani*}
\affiliation{Van der Waals-Zeeman Institute, Institute of Physics, University of Amsterdam, Amsterdam,The Netherlands}
\affiliation{Physical Chemistry and Soft Matter, Wageningen University \& Research,Stippeneng 4,Wageningen, 6708 WE Gelderland, The Netherlands}

\author{Nazanin Ghods*}
\affiliation{Institute of Process and Particle Engineering, Graz University of Technology, Inffeldgasse 13/III, 8010 Graz, Austria}

\author{Harkirat Singh*}
\affiliation{Department of Mechanical Engineering, Indian Institute of Technology Bombay, Mumbai 400076, India}

\author{Jing Wang}
\affiliation{Institute of Physics, Otto von Guericke University Magdeburg, D-39106 Magdeburg, Universit\"atsplatz 2, Germany}

\author{Ralf Stannarius}
\affiliation{Institute of Physics, Otto von Guericke University Magdeburg, D-39106 Magdeburg, Universit\"atsplatz 2, Germany}

\author{Stefan Radl}
\affiliation{Institute of Process and Particle Engineering, Graz University of Technology, Inffeldgasse 13/III, 8010 Graz, Austria}

\author{David L. Henann}
\affiliation{School of Engineering, Brown University, Providence, Rhode Island 02912, USA}

\author{Joshua A. Dijksman}
\affiliation{Van der Waals-Zeeman Institute, Institute of Physics, University of Amsterdam, Amsterdam,The Netherlands} 
 \affiliation{Physical Chemistry and Soft Matter, Wageningen University \& Research,Stippeneng 4,Wageningen, 6708 WE Gelderland, The Netherlands}

\def\thefootnote{*}\footnotetext{These authors contributed equally to this work}

\date{\today}

\begin{abstract}
Slowly sheared particulate media like sand and suspensions flow heterogeneously as they yield via shear bands, in which most strain accumulates. Understanding shear band localization from microscopics is still a major challenge. One class of so-called non-local theories identified that the width of the shearing zone $\xi$ should depend on the stress field, in particular through the local distance to the yield point of the material. We explicitly test this stress sensitivity picture by using a uniquely stress-tunable suspension while probing its flow behavior in a classic geometry in which shear bands are known to scale nontrivially with local stress: the Split-Bottom Shear Cell (SBSC).
The stress-tunable suspension is composed of mildly polydisperse soft, slippery hydrogel spheres submersed in water. We measure their flow profiles and rheology while controlling the confinement stress via both hydrostatic effects and compression. Unique for these soft particles is that we can probe flow fields under confining normal stresses that reach about 1\% of their elastic modulus. We determine the average angular velocity profiles in the quasi-static flow regime using Magnetic Resonance Imaging based particle image velocimetry (MRI-PIV) and discrete element method (DEM) simulations. We explicitly match a pressure-sensitive non-local granular fluidity (NGF) model to observed flow behavior. We find that shear bands for this type of suspension become extremely broad under the low confining stresses from the almost density-matched fluid particle mixture, while collapsing to a narrow shear zone under finite, externally imposed compression levels. The DEM and NGF results match the observations qualitatively, confirming the conjectured pressure sensitivity for suspensions and its role in the NGF model. Our results indicate that pressure sensitivity should be part of non-local flow rules to describe slow flows of granular media.   

\end{abstract}

\maketitle

\section{Introduction}
Granular materials are dense arrangements of particles that collectively can display solid-like or liquid-like behavior. They are crucial ingredients in applications as diverse as, e.g., geomechanics~\cite{ando2012grain}, food \cite{cataldo2009tdr}, battery assemblies \cite{gaitonde2016thermal}, pharmaceuticals \cite{wang2016predicting}, and ceramics \cite{curran1993micromechanical,duran2012sands}. Predicting how granular materials flow is thus important but challenging. Most studies on the flow behavior of granular materials have focused on simplified systems, composed of rigid, dry granular materials \cite{ma2018strain,gillemot2017shear,gdr2004dense,pouliquen2002friction}, where the effects of particle deformation and interstitial fluids are of minor importance for the particle dynamics. Much has been learned about the physics of such rigid, dry granular materials over the past several decades, yet it remains to be seen how the results from these works generalize or connect to other interesting ``granular systems''. In particular, the influence of stress gradients, particle softness, and the wide variety of (rate dependent) particle friction coefficients on the flow behavior has not been completely explored, posing a roadblock to understanding the generality of certain successful flow modeling approaches for the broader collection of particulate matter, such as, e.g., emulsions and foams~\cite{clark2020non}. Here, we use experiments and numerical simulations to show that a class of materials entirely different from the canonical glass bead mixtures, namely, dense, non-Brownian, nearly-frictionless, soft particle suspensions, can also be effectively modeled with a common class of non-local flow models, in which pressure sensitivity is built into a non-local length scale, governing the fluidization of the material. 

This flow modeling framework is applicable to extremely slow flows, where contacts are long lived and dominate the material response. This so-called ``quasi-static'' flow regime displays two important features: (1) rate independence of the driving stress and (2) broad shear bands, often modeled as non-local effects \cite{kamrin2012,bouzid2015PRL}. Non-locality in particle systems undergoing non-uniform steady flow is conjectured to arise from the idea of cooperativity, in which fluctuations at a point are induced by distant rearrangements, meaning that flow in one region can fluidize distant regions. Non-local continuum modeling has long been pursued to describe elastic and plastic material behavior \citep[e.g.,][]{aifantis2020concise} and, in recent years, a number of non-local models for dense granular flow have been developed that aim to account for cooperative effects \cite[e.g.][]{kamrin2012,bouzid2015non,pouliquen2009non}. Among these, the non-local granular fluidity (NGF) model \cite{kamrin2019non,henann2014continuum} has shown success in quantitative predictions of quasi-static, dense granular flow in certain flow geometries for stiff frictional glass beads \cite{kamrin2015nonlocal,henann2013predictive}. It should be noted that non-local features have also been attributed to ``granular temperature''  diffusion in dense flow extensions of kinetic theories~\cite{Berzi2024,Kamrin2024} and that a connection between the granular fluidity field on which the NGF model is based and the granular temperature has been established \cite{kim2020power}, suggesting an interpretation of the NGF model as a phenomenological simplification of granular kinetic theory. These discussions raise the question how one can design tests that distinguish the applicability and generality of specific modeling perspectives. Here we ask: how generally applicable is the NGF model to other dense granular systems beyond rigid, dry granular materials? 

\subsection{The role of pressure in soft particle suspensions}
To test the applicability of non-local modeling approaches for rigid, dry granular materials to more challenging granular systems, we probe the flow dynamics of a dense suspension of soft, frictionless hydrogel particles. Suspended hydrogel particles are an attractive model system for testing the pressure sensitivity of the NGF model, because they allow for investigating the relevance of three different but potentially important stress scales: the local pressure, the maximum hydrostatic pressure set by the particle-fluid density difference $\Delta\rho = \rho-\rho_f > 0$, and the particle Young modulus $E_p$. Such a suspension can be made by submersing almost buoyancy matched, swelled hydrogel spheres in water~\cite{workamp2019contact}. The submersion reduces the hydrostatic pressure gradient dramatically, which makes the suspension measurably sensitive to external pressure effects. In the present work, a small maximum hydrostatic pressure on the order of $1$\,Pa is outdone by an externally applied confining stress that we tune between $10-250$\,Pa. Regarding particle softness, the Young's modulus of the hydrogel is about $E_p=10$\,kPa. Therefore, while the maximum applied confining stress of 250\,Pa is much larger than the hydrostatic pressure, it remains sufficiently smaller than the particle Young's modulus, ensuring that both particle shape and pore fluid flow effects, induced by strong particle deformation, remain minimal. That said, the hydrogel particles are much softer than the glass beads ($E_p\approx 10$\,GPa) used in the canonical dry granular system, which are typically idealized as rigid, and prior works \citep{favier2017rheology,Singh2024} have demonstrated an effect of particle softness on rheological behavior. Overall, dense hydrogel particle suspensions enable rheological behavior to be probed under distinct combinations of these three stress scales compared to prior experimental investigations using, for example, glass beads. Finally, we note that poro- and viscoelastic effects in the hydrogel material itself can also affect the sensitivity of the hydrogel suspension to external pressure \citep{Shakya2024}, though they are not the focus of this work.

To create a stringent test for the continuum modeling, we perform experiments in the well-known split-bottom shear cell (SBSC) geometry, which creates non-trivial azimuthally symmetric two-dimensional flow fields. This geometry produces wide shear bands away from the sidewalls of the container \cite{dijksman2010granular}. We use Magnetic Resonance Imaging (MRI) as a tomographic technique to characterize the shape of the shear bands~\cite{Farmani2025}. MRI applied to hydrogels provides us with structural information from the bulk of the particle system at sub-mm resolution~\cite{wang2022characterization}. MRI can be combined with Particle Image Velocimetry (PIV) to obtain flow profiles and potentially even directly measure velocity fluctuations~\cite{Clarke2024}. The experimentally obtained profiles are compared with corresponding Discrete Element Method (DEM) simulations to confirm the observed flow fields and their pressure dependence. 

We then test the capability of the NGF model in predicting the experimentally and numerically observed bulk flows for cases both without applied pressure (unconfined) and applied pressure (confined). The NGF model incorporates pressure-dependence by way of its dependence on the stress field. Importantly, the NGF model utilizes a Drucker-Prager flow threshold under homogeneous flow conditions, in which the shear yield stress scales linearly with pressure. The static yield point in homogeneous flow also represents the stress state at which the length scale governing non-local, cooperative effects diverges, so the shear stress and pressure fields affect the features of shear bands in predictions of non-uniform steady flow. Moreover, the NGF model accounts for pressure in the rate dependence of dry granular flow through the dimensionless inertial number, but, as discussed in the next section (\ref{sec:introRD}), rate dependence arising due to particle inertia or fluid viscosity is expected to be negligible in the slow, quasi-static MRI experiments in the present work. One source of pressure sensitivity not accounted for in the NGF model is the role of particle stiffness $E_p$, quantified by the ratio of the pressure to the particle modulus as discussed above, due to the NGF model's prior focus on granular materials composed of nearly rigid particles. 
In spite of the absence of this physics in the NGF model, we find that the experimentally obtained slow flow profiles in the SBSC and their pressure sensitivity can be captured well with  NGF continuum modeling.

\subsection{Rate dependence}\label{sec:introRD}
In dry granular media, rate dependence may be quantified through the dimensionless inertial number~\cite{gdr2004dense} $I=\dot\gamma\sqrt{d^2\rho/P}$, in which $\dot{\gamma}$ is the local shear strain rate, $d$ the particle diameter, $\rho$ the particle-material density, and $P$ the local pressure. Submersing particles in a fluid introduces another rearrangement timescale that potentially affects the flow dynamics, so the inertial number is not the only non-dimensional form for the strain rate. Instead, the strain rate may be non-dimensionalized using the viscous time scale $\eta/P$, where $\eta$ is the viscosity of the fluid, resulting in the dimensionless viscous number $J = \eta\dot\gamma/P$ \citep{boyer2011unifying}. Then, the local rheological behavior depends on a viscoinertial combination of $I$ and $J$ \cite{amarsid2017viscoinertial,Guazzelli2024}. Using \textit{soft} particles for a granular flow study introduces yet another dimensionless timescale related to the modulus of the particles and the finite collision time such elasticity induces~\cite{CAMPBELL2002} and can contribute to complexities, such as for example non-monotonic flow behavior~\cite{workamp2019contact,Ness2025}. The MRI imaging method used in our experiments takes more than one minute to record one single single tomogram of a three-dimensional sample, so only quasi-static flows can be experimentally probed, in which grain inertia, elasticity-related dynamics, and interstitial fluid viscosity do not play a role. With the study parameters provided below, e.g., in Table~\ref{tab:table1}, and the assumption that $\dot{\gamma} \approx \omega$, an estimate of $I \approx 10^{-4}$ and $J \approx 10^{-3}$ support this view, even though these dimensionless numbers vary throughout the shear zone. For this reason, in the present work, we do not pursue a viscoinertial or elastic extension of the NGF model or its application to capturing the rate dependence of dense hydrogel particle suspensions.  We should further note that the hydrogel material of which the spheres are made is also known to be viscoelastic~\cite{dijksman2017refractive, Shakya2024, Shakya2025}, introducing additional poroelastic and viscoelastic timescales into the system. These modes are typically very slow: relaxation timescales can reach tens to thousands of seconds, which are of the order of time between single shear steps in the experiments and 3D scanning time scales. Therefore, these relaxation modes may be relevant, but they are not well understood enough~\cite{Shakya2024, Shakya2025} to fully appreciate their impact on the flow behavior of hydrogel suspensions. That said, we do explore the rheological behavior in the regime of slightly faster flows via flow dissipation measurements. We find that hydrogel suspensions at low to moderate levels of confinement exhibit Herschel-Bulkley-type rheological behavior, while at elevated confinement, a peculiar rate dependence is observed that is not captured by  rheological models with a rate independent static yield point. Comparison of the experimental rheological measurements with DEM methods highlights the need to incorporate other rate dependent effects in discrete flow modeling methods, for example via specific time-dependent contact laws.

\subsection{Structure of the paper}
This paper first presents an overview of experimental methods in Section~\ref{sec:expmethods}, including the MRI and rheometry techniques. We then discuss the implementation of the DEM methods in Section~\ref{sec:DEM}, after which the specifics of the NGF continuum model are laid out in Section~\ref{sec:NGF}, specialized for dense hydrogel particle suspensions. We then present a systematic comparison of the steady flow fields for all three methods in Section~\ref{sec:flowcomparison} to confirm their consistency, focusing first on unconfined flows with no applied pressure and then confined flows with increasing applied pressure. We then describe the rate dependent flow features of hydrogel suspensions in Section~\ref{rate_dependence}. We return to discuss the effect of stress fields on shear bands in quasi-static flows and the rate dependence of hydrogel suspensions in Section~\ref{sec:discussion} and close with some concluding remarks in Section~\ref{sec:conc}.

\begin{figure*}[!t]
\centering
\includegraphics[width=0.5\textwidth]{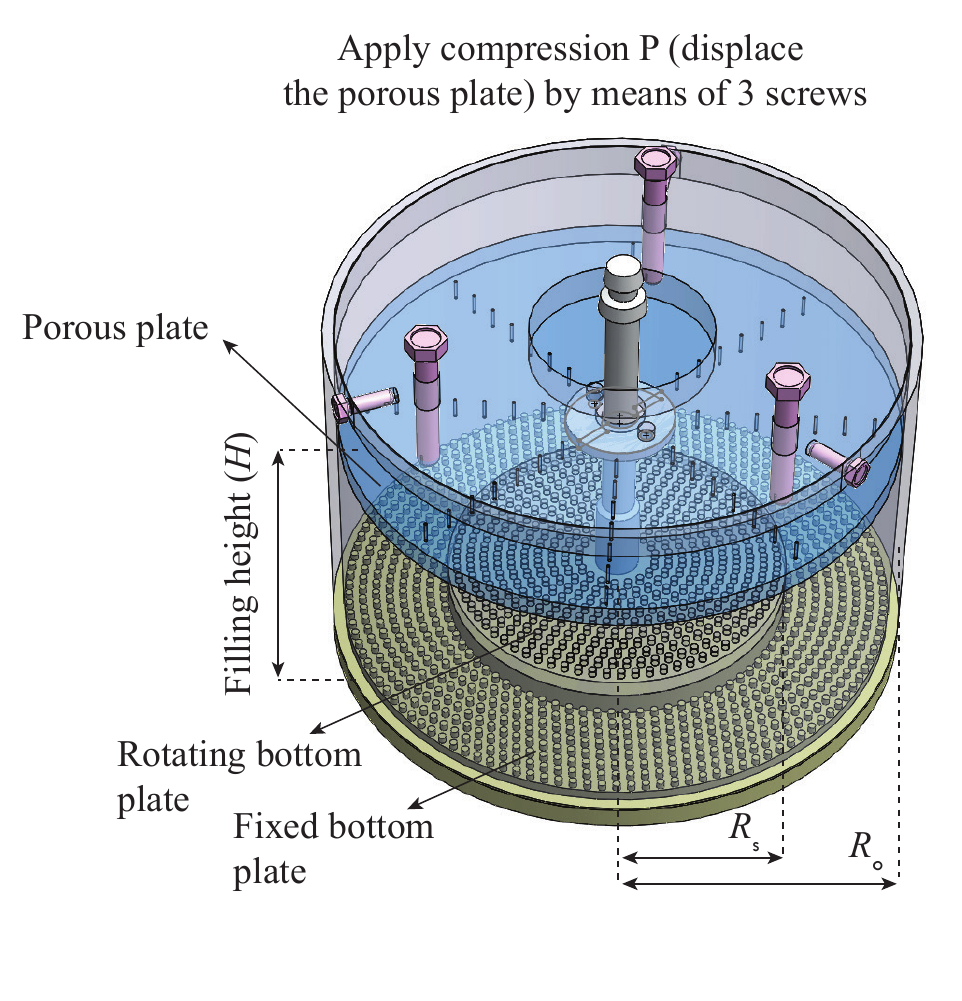}
\caption{Design of the SBSC for compression-shear measurements. Here, we use a porous plate to apply compression to the packing. We set the size of the holes in this plate to less than a particle diameter so that the particles cannot move through the holes. We also keep the gap between the central shaft and the porous plate less than a particle diameter to prevent particles from leaving the confined volume.}
\label{compsetup}
\end{figure*}

\section{Methods}
\subsection{Magnetic Resonance Imaging: flow experiments}\label{sec:expmethods} 
\textit{Geometry and sample ---} To study the shear behavior of dense collections of soft, frictionless hydrogel particles using MRI, we use $2$--$3$\,mm  hydrogel spheres that are swelled in water and fully submersed thereafter in water. Detailed information on the materials and methods used in the MRI measurements can also be found elsewhere~\cite{wang2022characterization,Farmani2025}. We designed and used a SBSC to shear/compress the hydrogel suspension with a constant rotation rate ($\omega_0$) and added confining pressure ($P$). The cell design is such that exactly the same geometry and boundary conditions can be used for both imaging and flow dissipation measurements with a rheometer. In the shear cell, a layer of material of depth $H$ is driven by the rotation of an inner disc of radius $R_S$, as shown in Fig.~\ref{compsetup}. This geometry is well studied \cite{woldhuis2009wide,fenistein2004universal,fenistein2003wide} and produces robust and wide shear zones for dry granular flows and suspensions~\cite{dijksman2010frictional} and for cohesive hydrogel spheres~\cite{Farmani2025}, in which the center position of the shear zone can be predicted by a dissipation minimization argument~\cite{unger2004shear, PhysRevLett.127.278003}. The ratio of the fill height to the radius of the rotating bottom disk, $H/R_S$, controls the geometry of the shear zone. As an additional parameter, we use the confining pressure $P$ exerted by the top plate on the granular phase of the suspension.

\begin{figure}[!t]
\centering
\includegraphics[width=0.9\textwidth]{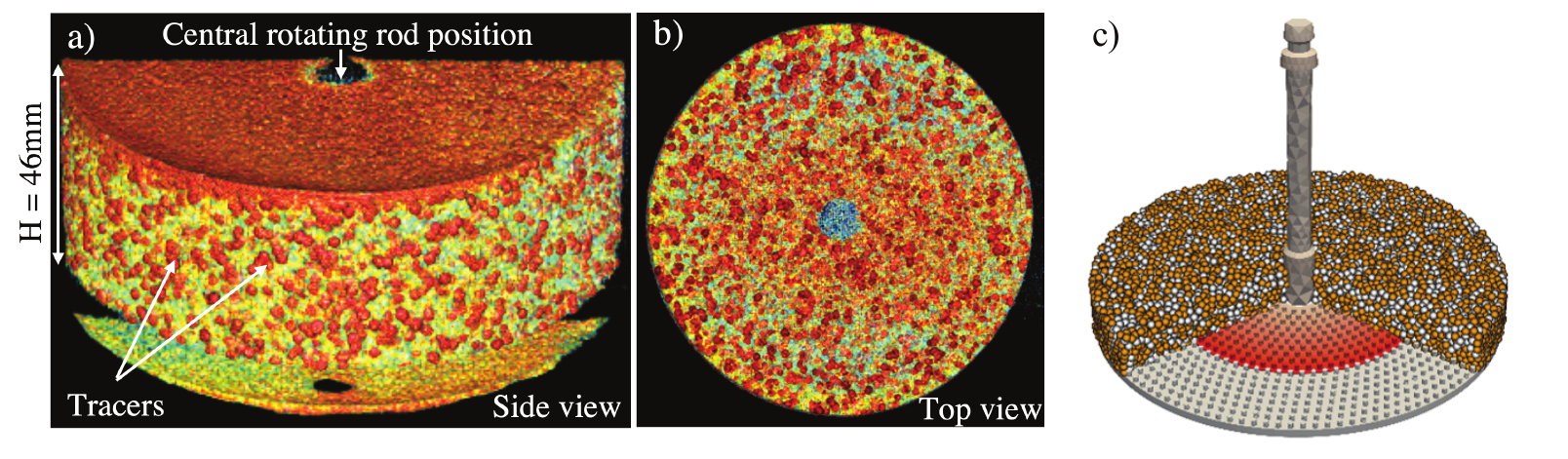}
\caption{Reconstructed tomograms: (a) side view and (b) top view of the hydrogel packing for a filling height of $H=46$\,mm. Red particles in the tomograms represent tracers that are hydrogels swollen in a dilute CuSO$_4$ solution, while yellow particles were swollen in pure distilled water. (c) Schematic of the DEM simulation geometry and distribution of the bi-dispersed packing (orange and grey represent particles with diameters of 2.8~mm and  2.0 mm, respectively) for $H=15$\,mm.}
\label{tomograms}
\end{figure}

\textit{Experimental protocols ---} The SBSC experiments on submersed hydrogel particles were executed via two different protocols. In procedure 1, we performed stepwise shear, consisting of five parts. The stepwise shear protocol is necessary since the imaging instrument is a medical MRI scanner and can only image the three dimensional structure of static objects. To image a quasi-static flow, we therefore record the deformation of the material step-by-step, as is customary for such extremely slow flows~\cite{fenistein2003wide, Sakaie2008, Farmani2025}. In part 1 of the experiment, the shear cell was filled with hydrogels spheres up to a certain height $H$; in part 2, the shear cell was then put inside an MRI scanner; and in part 3, we applied a slow, manual, 10\textdegree\, rotation ($\omega_0\approx0.025\, {\rm rad/s}$) using the central rotating shaft of the cell. In part 4, three dimensional (3D) tomograms were taken after each stepwise rotation (Fig.~\ref{tomograms}), and in part 5, this procedure was repeated for different filling heights $H$. We performed a total of 7 shear steps to limit experimental time per run. The justification for limiting our experiments to 7 shear steps is that we have previously carried out a similar experiment over a full rotation with 19 shear steps, in which it was found that 2 steps are sufficient to reach a fully developed steady flow field. For our analysis, the first two shear steps are excluded, as the flow field has not yet reached a steady state. All original data and analysis scripts can be found in a Zenodo repository~\cite{farmani_2023_10729929}.

\begin{figure}[!t]
\centering
\includegraphics[width=0.8\textwidth]{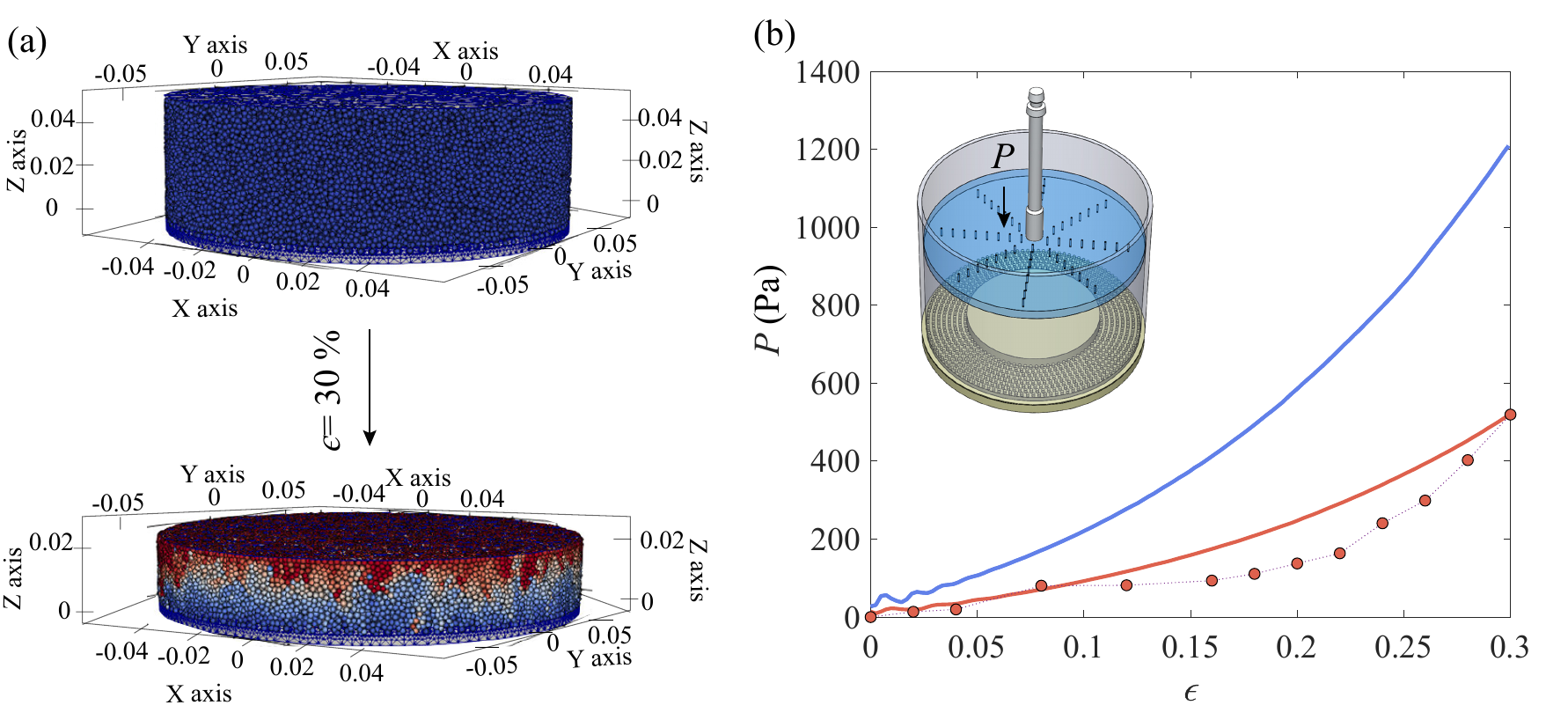}
\caption{Calibration of the particle Young's modulus $E_p$ based on measurements of the applied pressure exerted on the hydrogel suspension for a filling height of $H=50$\,mm. (a) DEM simulation of the packing before and after applying a uniaxial strain of 30\%. The coordinate directions $x,y,z$ used in this work are indicated. All dimensions are in meters. (b) Normal stress ($P$) as a function of applied uniaxial strain ($\epsilon$) on the porous plate, experiment \protect\tikzmycircle[fill=Red1]{2pt}, DEM simulation with $E_p = 23.3$\,kPa \cite{brodu2015spanning} \protect\stressline, and DEM simulation results with a calibrated value of $E_p$ = 10 kPa \protect\stresslinee.}
\label{Youngs}
\end{figure}

For procedure 2, we sheared stepwise under compression. After part 1 from procedure 1, we put a porous plate on top of the packing and compressed the packing by displacing the porous plate via set screws; see Fig.~\ref{compsetup}. Then we continued the same above-mentioned shear steps for 6 compression levels. The static stress corresponding to each step-wise 2~mm compression was calibrated separately in the rheometer by compressing a packing of $H=50$~mm in the absence of the central shaft and the rotating plate (symbols in Fig.~\ref{Youngs}(b)). This standalone compression setup was replicated numerically in DEM simulations for verification and calibration purposes, with satisfactory agreement given the limitation that the DEM does not include many-body effects~\cite{Brodu2015}. 

\textit{Image analysis ---} MRI measurements provide time-averaged azimuthal velocity fields $v_\theta(r,z)$ with high spatial resolution by PIV, with $z$ the distance from the top surface. These profiles are then used to calculate the angular velocity fields $\omega(r,z)=v_\theta(r,z)/r$. Non-dimensionalizing $\omega(r,z)$ by the angular velocity of the bottom plate $\omega_0$ gives fields that vary between zero and one. We determined the average angular velocity via a standard autocorrelation analysis of azimuthal image intensity profiles in consecutive image stacks~\cite{workamp2019contact}. This approach allowed us to compare MRI data with DEM simulations and NGF model predictions for both unconfined and confined flows. To extract velocity profiles, the 3D tomograms from MRI were divided into rings with constant depth $z$ below the free surface and constant distance $r$ from the central axis. We used the auto cross-correlation method on said domains to extract the displacement field, which may be used to calculate the angular velocity ($\omega$) at certain $z$ and $r$. We note that at smaller $r$, the number of pixels we can use for autocorrelation is smaller, hence the results are more noisy. Data were averaged over 5 shear steps for each $H$, removing an initial transient of 2 shear steps.  The angular velocity imposed by the moving boundary disk $\omega_0$ can be independently measured from the same cross-correlation analysis of displacements of tracer hydrogel beads glued to the underside of the disk, which are also imaged in the same tomograms at each shear step.

\textit{Flow dissipation ---} To investigate the flow dissipation behavior of the hydrogel suspension, we conducted two sets of experiments: one for unconfined suspensions and another under confined conditions. All experiments were performed with an Anton Paar MCR 301 rheometer equipped with a custom-designed plate mounted on the rheometer shaft. In both cases, we measured the torque $T$ during a rate ramp ranging from $10^{-3}$ to $10^{1}$~rad/s. For the unconfined case, we varied the packing height and recorded the corresponding torque at each shear rate. For the confined case, the packing height was kept constant while the confining pressure was adjusted by placing a top plate on the packing and displacing it to different positions. To identify the rate-independent regime, we also performed tests at fixed shear rates of 0.001 and 0.01~rad/s, measuring the average torque under steady conditions. To enable direct comparison with our DEM simulations, which also provide torque vs. rate data, we performed torque calibration using dry glass beads (see Fig.~\ref{figure20} in Appendix~\ref{torque}). For calibration, we measured the torque across four different filling heights: $H = 8$, 10, 15, and 18~mm.

\subsection{Discrete Element Method}\label{sec:DEM} 
To gain deeper insight on how the micro-properties of the submersed hydrogel suspension can affect the flow behavior, the experiments were replicated by discrete element method-based simulations (DEM). The discrete element method (DEM) is a widely utilized approach to investigate the flow characteristics of granular materials. This method involves treating each particle as a separate element and solving Newton's equations to determine the translational and angular momentum. In this work, the particles were modeled as spherical in shape, possessing elastic properties, and interacting with one another through Hertz's contact law, so that the viscoelastic interaction force between particles $i$ and $j$ is given by
\begin{equation}\label{DEMforce}
\mathbf{F}_{ij}=\mathbf{F}_{{\rm n}ij}+\mathbf{F}_{{\rm t}ij}
\end{equation}
with
\begin{equation}
\mathbf{F}_{{\rm n}ij}=\sqrt{\delta_{ij}R_{\rm eff}}\left[k_{\rm n}\delta_{ij}-\gamma_{\rm n}m_{\rm eff\ }\mathbf{v}_{{\rm n}ij}\right]\quad \text{and}\quad \mathbf{F}_{{\rm t}ij}=\sqrt{\delta_{ij}R_{\rm eff}}\left[{-k}_{\rm t}\mathbf{u}_{{\rm t}ij}-\gamma_{\rm t}m_{\rm eff}\mathbf{v}_{{\rm t}ij}\right].
\end{equation}
Here, $\delta_{ij}$ is the normal overlap of particles $i$ and $j$, $R_{\rm eff}=R_iR_j/(R_i+R_j)$ is the effective radius of two contacting particles $i$ and $j$ with radii $R_i$ and $R_j$, $k_n$ and $k_t$ are the normal and tangential stiffness constants, $\gamma_n$ and $\gamma_t$ are the viscous damping constants in the normal and tangential directions, respectively, $m_{\rm eff} = m_i m_j /(m_i+m_j)$ is the effective mass for particle masses $m_i$ and $m_j$, $\mathbf{v}_{{\rm n}ij}$ and $\mathbf{v}_{{\rm t}ij}$ are relative particle velocities in the normal and tangential directions, and $\mathbf{u}_{{\rm t}ij}$ is the elastic shear displacement, which is bounded to fulfill the static yield criterion, $|\mathbf{F}_{{\rm t}ij}|\le\mu_{p-p}|\mathbf{F}_{{\rm n}ij}|$, where $\mu_{p-p}$ is the particle friction coefficient. Denoting Young's modulus and Poisson's ratio of particle $i$ as $E_i$ and $\nu_i$, respectively, the normal and tangential stiffness constants are $k_n=(4/3)E^\ast$ and $k_t=8G^\ast$, where ${E^\ast}=\left[{\left(1-\nu_i^2\right)}/{E_i}+{\left(1-\nu_j^2\right)}/{E_j}\right]^{-1}$ and ${G^\ast}=\left[{2\left(2-\nu_i\right)\left(1+\nu_i\right)}/{E_i}+{2\left(2-\nu_j\right)\left(1+\nu_j\right)}/{E_j}\right]^{-1}$. The normal and tangential viscous damping constants are $\gamma_n=-2\beta\sqrt{5{S_n}/6{m_{\rm eff}}}$ and $\gamma_t=-2\beta\sqrt{5{S_t}/6{m_{\rm eff}}}$, where $\beta={\ln{\left(e_{p-p}\right)}}/({{\ln}^2{\left(e_{p-p}\right)}+\pi^2})$, $e_{p-p}$ is the particle coefficient of restitution, $S_n={2E^\ast}/{\sqrt{R_{\rm eff}\delta_{ij}}}$, and $S_t={8G^\ast}/{\sqrt{R_{\rm eff}\delta_{ij}}}$. Further details on the contact law can be found elsewhere in the literature \cite{ghods2022}.

\textit{Hydrodynamic forces ---} To study the flow behavior of submersed hydrogel suspensions, besides the viscoelastic interaction force in \eqref{DEMforce}, two other forces acting on each particle have to be taken into account. These are the buoyancy ($\mathbf{F}_{\rm B}$) and viscous lubrication forces ($\mathbf{F}_{\rm V}$). The buoyancy force cannot be neglected as it counteracts the gravitational force, especially in the case of hydrogel particles that are almost perfectly density-matched with the fluid, and is given by 
\begin{equation}
\mathbf{F}_{\rm B}={-\rho_f}\mathbf{G} V_{p},
\end{equation}
where $\rho_f$ is the fluid density, $\mathbf{G}$ is the gravitational acceleration vector, and $V_p$ is the particle volume. The hydrogel particles are also at all times fully submersed, so there are no capillary bridges~\cite{Farmani2025}, but the viscous environment does affect the forces on the particles. An important hydrodynamic viscous force, known as the lubrication force, arises from a radial pressure within the interstitial fluid, which is squeezed from the space between adjacent particles. Previous work has examined elastohydrodynamic collisions between spherical particles \cite{davis1986elastohydrodynamic}. Here, we used the analytical approximation based on a Hertzian-like profile for the elastic deformation of two spheres, proposed by \citet{lian1996elastohydrodynamic}, to incorporate the lubrication force into DEM. The normal and tangential components are given by the models of \citet{adams1985cohesive} and \citet{goldman1967slow}, respectively:
\begin{equation}
\mathbf{F}_{{\rm V,n}ij}=-6\pi\eta\ R_{\rm eff}\mathbf{v}_{{\rm n}ij}/D^*\quad\text{and}\quad \mathbf{F}_{{\rm V,t}ij}=-6\pi\eta\ R_{\rm eff}\mathbf{v}_{{\rm t}ij}\left(\frac{8}{15}\ln\frac{1}{D^*}+0.9588\right),
\end{equation}
where $\eta$ is the dynamic viscosity of the fluid, and $D^*$ is defined as $\max(D_{\min}/R_{\rm eff},D/R_{\rm eff})$ and is the dimensionless surface-to-surface distance ($D$) between two particles or between a particle and the wall. The parameter $D_{\min}$ is the minimum distance, which may be interpreted physically as the sum of the absolute roughnesses of two touching surfaces, introduced to prevent the value of the lubrication force from becoming too large. Here, $D_{\min}$ is taken to be $0.01R_{\rm eff}$. The lubrication force becomes active as soon as the distance between two surfaces is below a threshold maximum distance, which here is set to be $0.1(R_{i}+R_{j})$.

\textit{Geometry and sample parameters ---} The geometry used in the DEM simulations is identical to the experimental setup with 1:1 scaling and with the same roughness of the bottom plates (Fig.~\ref{tomograms}). This is an advantage since boundary roughness is often created by gluing irregularly shaped particles to the bottom plates, which cannot be exactly reproduced numerically \cite{singh2014effect}. To prevent the formation of crystalline structure in the granular media \cite{jing2017effect}, a 50:50 (mass-based) binary packing with a diameter ratio of 1:1.4 is used, where the diameter of smaller particles is 2~mm. Moreover, test simulations with common neighbor analysis \cite{tsuzuki2007structural} were conducted to ensure that no crystallized pattern is generated during the shearing process.  

\begin{table}[!t]
    \centering
    \caption{DEM model parameters.}
    \begin{tabular}{|l|l|}
    \hline
       Parameter  & Value \\
       \hline
         Particle density ($\rho$) & 999 kg/m$^3$\\
         Particle-Particle sliding friction ($\mu_{p-p}$) & 0.01\\
         Particle-Wall sliding friction ($\mu_{p-w}$) & 0.01\\
         Particle Young's Modulus ($E_{p}$) & 10 kPa\\
         Wall Young's Modulus ($E_w$) & 10 kPa\\
         Coefficient of restitution ($e_{p-p}= e_{p-w}$) & 0.5\\
         Particle Poisson's ratio ($\nu_p$) & 0.5\\
         Wall Poisson's ratio ($\nu_w$) & 0.3\\
         Gravity ($G$) & 9.81 m/s$^{2}$\\
         Water density ($\rho_{f}$) & 997 kg/m$^{3}$\\
         Dynamic viscosity of water ($\eta$)	 & 1.0016\,mPa$\cdot$s\\
         Rotation rate ($\omega_0$)	 & 0.02\,rad/s\\
         \hline
    \end{tabular}
    \label{tab:table1}
\end{table}

The particle properties and DEM parameters used are shown in Table~\ref{tab:table1}. The particle properties were selected from previous work~\cite{ghods2022}. The volume of the hydrogel particles is 99.9\% water, and, since it is not possible to completely isolate the particles from the interstitial liquid (i.e., they cannot reach a dry state), it is challenging to measure the effective density difference between water and the hydrogel material to the necessary level of accuracy. Therefore, we used DEM to infer this difference from a calibration of DEM results against rheology tests provided by the torque-shear rate data (Fig.~\ref{figure8}). The total torque on the body exerted by the particles is computed by summing the torque on each mesh element of the split disk in reference to the rotating point, which is the center of the split disk. The torque measurement was verified for glass beads (see Appendix~\ref{torque}). The calibration resulted in a relative density of 2 kg/m$^{3}$, i.e., a density mismatch between particle and fluid; see Table~\ref{tab:table1}. The particle Young's modulus ($E_p$) was calibrated by simulating the stand-alone compression of a packing of hydrogel particles with a filling height of 50\,mm in the same setup as the corresponding experiment (Fig.~\ref{Youngs}). We also verified that the wall's Young's modulus does not affect the flow profiles. Therefore, it was set to that of the particles in order to decrease computational efforts. 

\textit{Shear protocol ---} The bottom plate was rotated with the same angular velocity as in the experimental protocols, but, unlike in our experiments, the shearing process in the DEM simulations was performed continuously, not stepwise. The DEM data corresponding to the strain equivalent of the first two experimental shear steps were ignored in order to reach a fully developed steady state and avoid reordering effects. Then, DEM data were collected for subsequent rotation of the split-bottom disk. To quantify the average flow field of the particles in the SBSC, the angular velocity $\omega=v_{\theta}/r$ is calculated for each particle, where $v_{\theta}$ is the particle velocity in the azimuthal direction, and $r$ is the distance from the vertical axis along the center of the cell. The angular velocity is normalized by the angular velocity of the split-bottom disk ($\omega_0$). To retrieve representative, time-averaged continuum fields of the strain rate (and later also the stress components), we assumed that there is invariance along the azimuthal direction. Therefore, the velocities can be averaged both spatially and temporally in ring-shaped subregions of the SBSC. The ring elements have the height of the average particle diameter and form 20 bins in radial direction.

\subsection{Continuum model: Non-local Granular Fluidity}\label{sec:NGF}
The non-local granular fluidity (NGF) model has shown success in quantitatively predicting granular rheology in varied geometries \cite{kamrin2012,henann2013predictive,liu2017}, including the split-bottom flow configuration \cite{henann2013predictive,li2020nonlocal}. However, in previous studies on split-bottom flow, the model has primarily been tested against surface flow field data; here, we enable comparison of NGF model predictions with experimental bulk flow measurements. We can thus use this test bed to assess the capability of the NGF model in predicting bulk flows of soft hydrogels both without and with confining pressure. We proceed with an overview of the NGF methodology, including the process utilized to determine the parameters of the NGF model for a dense system of soft hydrogel particles. Regarding notation, we use component notation, in which the components of vectors, $\boldsymbol{v}$, and tensors, $\boldsymbol{\sigma}$, relative to a set of Cartesian basis vectors $\{\boldsymbol{e}_i|i=1,2,3\}$ are denoted by $v_i$ and $\sigma_{ij}$, respectively, and the Einstein summation convention is used.

\textit{Definition of the model parameters ---} The non-local granular fluidity (NGF) model builds upon the well-known local inertial, or $\mu(I)$, rheology \cite{gdr2004dense,dacruz2005,jop2005,jop2006,kamrin2010nonlinear}, which relates the local stress state at a point to the local state of the strain rate through a one-to-one constitutive relation. The stress ratio, denoted as $\mu$, is defined as $\mu=\tau/P$, where $\tau = \sqrt{\sigma'_{ij}\sigma'_{ij}/2}$ is the equivalent shear stress, $P=-\sigma_{kk}/3$ is the pressure, $\sigma_{ij}$ is the symmetric Cauchy stress tensor, and $\sigma'_{ij} = \sigma_{ij} + P\delta_{ij}$ is the deviatoric part of the stress. Here, $\delta_{ij}$ denotes the Kronecker delta, while in Section~\ref{sec:DEM}, it denotes the normal overlap of two particles. The non-dimensional form of the local strain rate is given by the inertial number $I=\dot\gamma\sqrt{d^2\rho/P}$, where $\dot{\gamma} = \sqrt{2 D_{ij} D_{ij}}$ is the equivalent shear strain rate, $D_{ij} = (\partial v_i/\partial x_j + \partial v_j/\partial x_i)/2$ is the strain-rate tensor, $v_i$ is the velocity vector, $\rho$ is the particle-material density, and $d$ is the average particle diameter. It is presumed that, after any confining pressure has been applied, steady flow proceeds at constant volume, so that the steady-state velocity field is divergence-free $\partial v_k/\partial x_k=0$, and the strain-rate tensor is deviatoric $D_{kk}=0$. The local inertial rheology then relates $\mu$ and $I$ through an empirical constitutive relation, which we denote as $\mu_{\rm loc}(I)$. Here, we used a phenomenological power-law form for the local rheology, which has been widely used in the literature \cite{degiuli2015,degiuli2016phase,srivastava2021}, especially for granular systems with a low particle-particle friction coefficient, and is given as
\begin{equation}\label{local_rheo}
\mu_{\rm loc}(I) = \mu_s + b\,I^{\rm \alpha},
\end{equation}
where $\mu_s$ is the critical stress ratio below which the local rheology predicts \textit{no flow}, and $\{b,\alpha\}$ are dimensionless parameters that characterize the rate dependence of the local rheology, so that \eqref{local_rheo} has three adjustable parameters $\{\mu_s, b, \alpha\}$. We have confirmed that the parameter $\mu_s$ is the only aspect of \eqref{local_rheo} that affects the model predictions in the quasi-static regime and that the functional nature of the dependence on $I$ and the parameters $\{b,\alpha\}$ have little effect. To capture the rate-dependent results, as, for example, described in Section~\ref{rate_dependence}, a viscoinertial generalization of \eqref{local_rheo}, incorporating the dimensionless viscous number $J=\eta\dot\gamma/P$, and an accompanying generalization of the NGF model is required, but this is beyond the scope of the present work. 

\textit{The phase field approach ---}
A local rheological modeling approach can capture the features of homogeneous shear flow. Moreover, when the local inertial rheology is applied to more complicated, inhomogeneous flows, such as dense heap flow, it can accurately predict rapidly flowing regions \cite{jop2005,jop2006,kamrin2010nonlinear} but predicts frozen zones in quasi-static regions where $\mu<\mu_{\rm s}$, which is contrary to experimental observations wherein decaying creeping zones are observed. Quasi-static, creeping regions are also observed in other flow configurations, such as annular shear flow \cite{losert2000particle,mueth2000signatures,koval2009annular} and SBSC flow \cite{fenistein2003wide,fenistein2004universal}, which local modeling approaches are unable to capture. To address this shortcoming, the NGF model was developed \cite{kamrin2012, henann2013predictive} wherein a positive, scalar state variable (or ``phase field''), called the granular fluidity $g$, is introduced, which relates the Cauchy stress deviator and the strain-rate tensor through $\sigma_{ij}' =  2(P/g)D_{ij}$, where we have made the idealization that the stress deviator and strain-rate tensors are codirectional \cite{rycroft2009assessing}. Taking the magnitude of this expression implies the scalar expression $\dot\gamma = g\mu$. Moreover, the granular fluidity field is governed by the following partial differential equation (PDE): 
\begin{equation}\label{nonlocal_seg}
g = g_{\rm loc}(\mu,P) + \xi^2(\mu) \frac{\partial^2 g}{\partial x_k \partial x_k },
\end{equation} 
where $g_{\rm loc}(\mu,P)$ is the local fluidity function and $\xi(\mu)$ is the stress-dependent cooperativity length. It is important to note that the PDE \eqref{nonlocal_seg} is the steady-state approximation of a dynamical PDE, which we discuss in further detail in Appendix~\ref{app:NGFss}. In this work, the local fluidity function is defined in a manner consistent with the power-law functional form \eqref{local_rheo} and the definition of the inertial number, and is given as 
\begin{equation}\label{eq:local_fluidity}
g_{\rm loc}(\mu,P)=  \left\{\begin{array}{cl}\sqrt{\dfrac{P}{{d}^2\rho}} \,\dfrac{1}{\mu}  \left(\dfrac{\mu-\mu_s}{b}\right)^{1/\alpha}  &{\rm if}\ \mu>\mu_s,\\[16pt]
0 &{\rm if}\ \mu\le\mu_s.\end{array}\right.
\end{equation}
The functional form for the cooperativity length $\xi(\mu)$ that is associated with the power-law local rheology (see Appendix~\ref{app:NGFss}) is
\begin{equation}\label{eq:cooperativity_length}
\xi(\mu)
=  \left\{\begin{array}{cl}\dfrac{Ad}{\sqrt{\alpha(\mu - \mu_s)}}  &{\rm if}\ \mu>\mu_s,\\[16pt]
\dfrac{Ad}{\sqrt{\mu_s - \mu}} &{\rm if}\ \mu\le\mu_s,\end{array}\right.
\end{equation}
where $A$ is the non-local amplitude, which is a dimensionless parameter that determines the extent to which non-local effects are spread. Thus, the non-local granular fluidity model is completely described by the parameter set $\{\mu_s, b, \alpha, A\}$. We estimated the parameters by fitting the bulk flow fields predicted by the model to the experimental MRI data in the absence of confining pressure, as described in the section below. The system of governing equations is solved for SBSC flow using the finite-element-based approach described in \citet{henann2016finite}, and two-dimensional, generalized axisymmetric numerical simulations of SBSC flow are performed using a similar finite-element setup and boundary conditions (both mechanical and fluidity) as described in prior work \citep[e.g.,][]{li2020nonlocal}.\\

\begin{figure}[!t]
\centering
\includegraphics[width=0.6\textwidth]{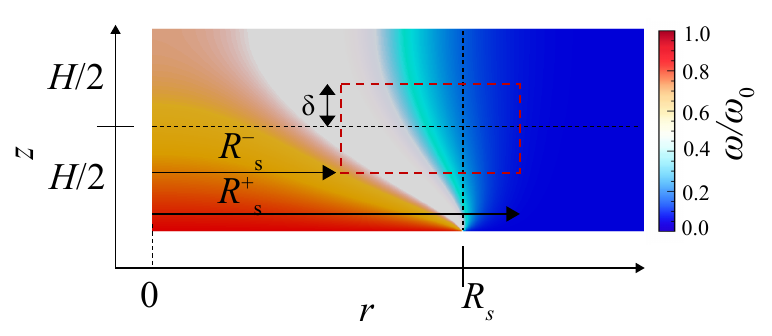}
\caption{Illustration of the region of interest (ROI) for a representative case of SBSC flow utilized in parameter estimation for the NGF model.}
\label{appendix_fitting}
\end{figure}

\textit{Parameter estimation for hydrogel particles ---} The version of the NGF model utilizing the power-law local rheology involves four dimensionless parameters $\{\mu_s, b , \alpha, A\}$. Past works have determined parameters, such as $\mu_s$ and $A$, for stiff, frictional spheres. The dense systems of submersed hydrogel spheres considered in the present work are different in the following two ways. Firstly, the contacts are nearly frictionless, and secondly, the hydrogels particles are soft. Therefore, a new set of parameters must be estimated. As a starting point for the local rheological parameters, we considered a parameter set similar to that determined by \citet{srivastava2021} for stiff, frictionless spheres using DEM simulations, $\{\mu_s=0.1, b = 0.52 , \alpha = 0.37\}$. We note that $\mu_s\approx0.1$ is consistent with other works on frictionless particles \citep{peyneau2008,Kawasaki2015}. One could then sweep over all four parameters and determine a best-fit set by fitting to experimental flow-field measurements; however, as discussed above, we have confirmed that the two parameters $b$ and $\alpha$ have little effect on model predictions in the quasi-static regime. Therefore, consistent with our focus on quasi-static flows, we keep them fixed at $\{b=0.52, \alpha=0.37\}$ and only fit the two parameters $\{\mu_s,A\}$ to selected experimental flow-field measurements in the absence of confining pressure. Keeping $b$ and $\alpha$ fixed, we vary $\mu_s$ in the range $[0.05,0.35]$ by increments of $0.05$, and $A$ in the range $[0.1,0.7]$ by increments of $0.1$, thus considering 49 combinations of $\mu_s$ and $A$. For each combination of parameters, we consider SBSC flow in absence of confining pressure for four different fill heights: $H = 20, 28, 45,$ and 50\,mm. Next, we choose a two-dimensional region of interest (ROI), which spans the area $R_S^- \le r \le R_S^+$ and $(H/2 -\delta)\le z \le (H/2 + \delta)$ as illustrated in Fig.~\ref{appendix_fitting} for a representative case of SBSC flow. The ROI spans both sides of the split radius and the mid-plane and prioritizes capturing the salient features of the shear band in this flow configuration. Throughout, we choose  $R_S^- = 0.7 R_S$, $R_S^+ = 1.1 \, R_S$, and $\delta=7$\,mm, and we have verified that slight changes in the dimensions of the ROI do not have a strong influence on the best-fitted parameters. Finally, we evaluate the root mean square (RMS) value of the difference between the experimental angular velocity field and the angular velocity field predicted by the continuum model over the ROI for a given set of parameters as 
$$\sqrt{\dfrac{1}{2\delta(R_S^+-R_S^-)}\int_{H/2-\delta}^{H/2+\delta}\int_{R_S^-}^{R_S^+}(\omega_{\rm exp} - \omega_{\rm NGF} )^2\,drdz}.
$$  We repeated this procedure for the four fill heights listed above and average the RMS values to obtain a single RMS value for each combination of material parameters. Based on the minimum RMS value for all combinations, we obtain the following set of parameters for the dense system of submersed hydrogel spheres: $\{\mu_s=0.1, b = 0.52 , \alpha = 0.37, A=0.4\}$. We note that the value of $\mu_s$ determined in this way is almost the same as the value determined by \citet{srivastava2021} for stiff, frictionless spheres, and the value of $A$ is similar to the value determined by \citet{zhang2017microscopic} for stiff, frictional spheres. As the calibration is done on unconfined flows, it is to be expected that the parameters for soft, frictionless spheres are similar to those of stiff, frictionless spheres. For comparison, we also note that the parameters determined for moderately stiff, frictional, two-dimensional disk flows~\cite{tang2018nonlocal} are $\mu_s=0.26$ and $A=0.4$ (keeping $b=1$ and $\alpha=1$ fixed). Since the dimensionality and stiffness are different for this system, one should not expect perfect agreement, but the value of $A$ is surprisingly similar.

\begin{figure*}[!t]
\centering
\includegraphics[width=0.8\textwidth]{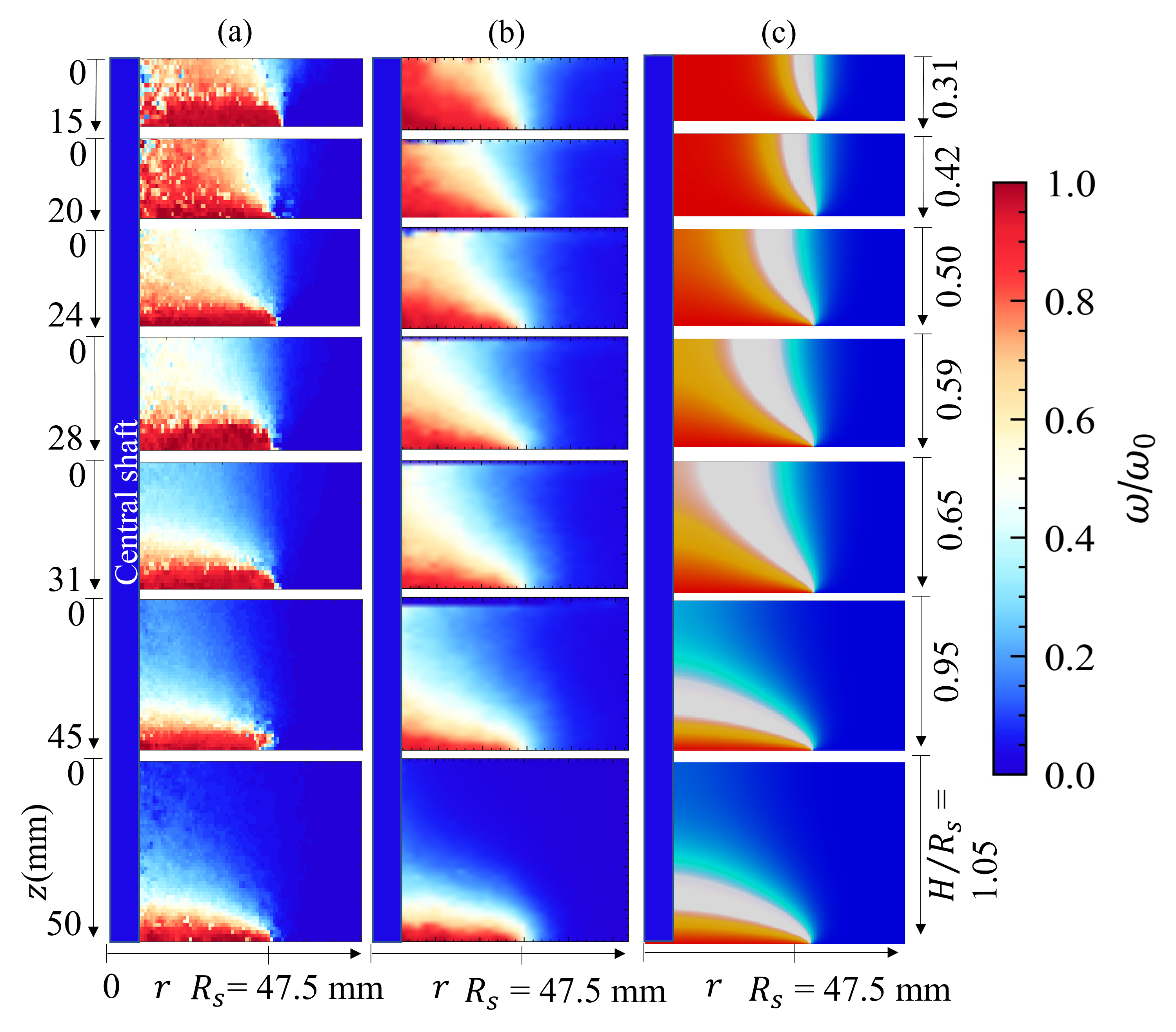}
\caption{Non-dimensional, steady-state angular velocity fields for dense hydrogel suspensions in a SBSC with no confining pressure $P=0$ and filling heights of $H =15$, 20, 24, 28, 31, 45, and 50\,mm from (a) MRI measurements, (b) DEM simulations, and (c) NGF continuum modeling. The central shaft is shown as a blue rectangle to the left of the 2D profiles. Static zone (\textcolor{blue}{\SquareSolid}), rotating zone (\textcolor{Red2d}{\SquareSolid}), and shear band (\textcolor{White2d}{\SquareSolid}).}
\label{xxxflowprof}
\end{figure*}

\section{Bulk flow profile comparisons for the MRI, DEM, NGF approaches}\label{sec:flowcomparison} 

Having introduced the methods for the experimental MRI, DEM simulation, and continuum-level NGF model approaches, we now explain how we systematically explored their flow profiles for different $H/R_S$ and in different confining stress situations.

\textit{Unconfined hydrogel packing flow behavior ---} We show the angular velocity fields in a dense hydrogel suspension from the MRI measurements for seven fill heights in Figure~\ref{xxxflowprof}a. Similarly, we used DEM simulations and the NGF model to produce flow profiles in the same domains, as shown in Figs.~\ref{xxxflowprof}b and c. In Figs.~\ref{xxxflowprof}a-c, we observe the shear band marked as a white region between the rotating zone (red) and the static zone (blue). The shear band, in general, is broader than the typical shear band in dense flows of frictional particles for all $H$ \cite{dijksman2010frictional}. A narrower shear band is observed at low $H$. (See Appendix~\ref{app:shearband} for further characterization of the shear band features.) The shear band reaches the surface up to $H \approx 0.59 R_S$, but then, as $H$ is increased further, the system moves towards a dome structure. It is noteworthy to observe that a bulk-confined dome does not appear, and even at $H \approx R_S$, there are small rotations close to the surface, although with reduced rate. We confirm observing wide shear bands or ``shear zones'' using DEM simulations (Fig.~\ref{xxxflowprof}b) and NGF continuum modeling (Fig.~\ref{xxxflowprof}c).

\begin{figure}[!tbp]
\centering
\includegraphics[width=0.45\textwidth]{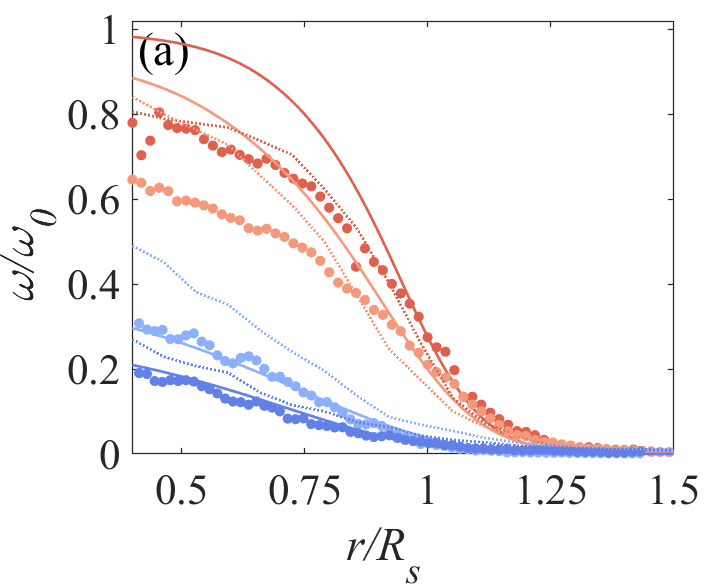}
\includegraphics[width=0.45\textwidth]{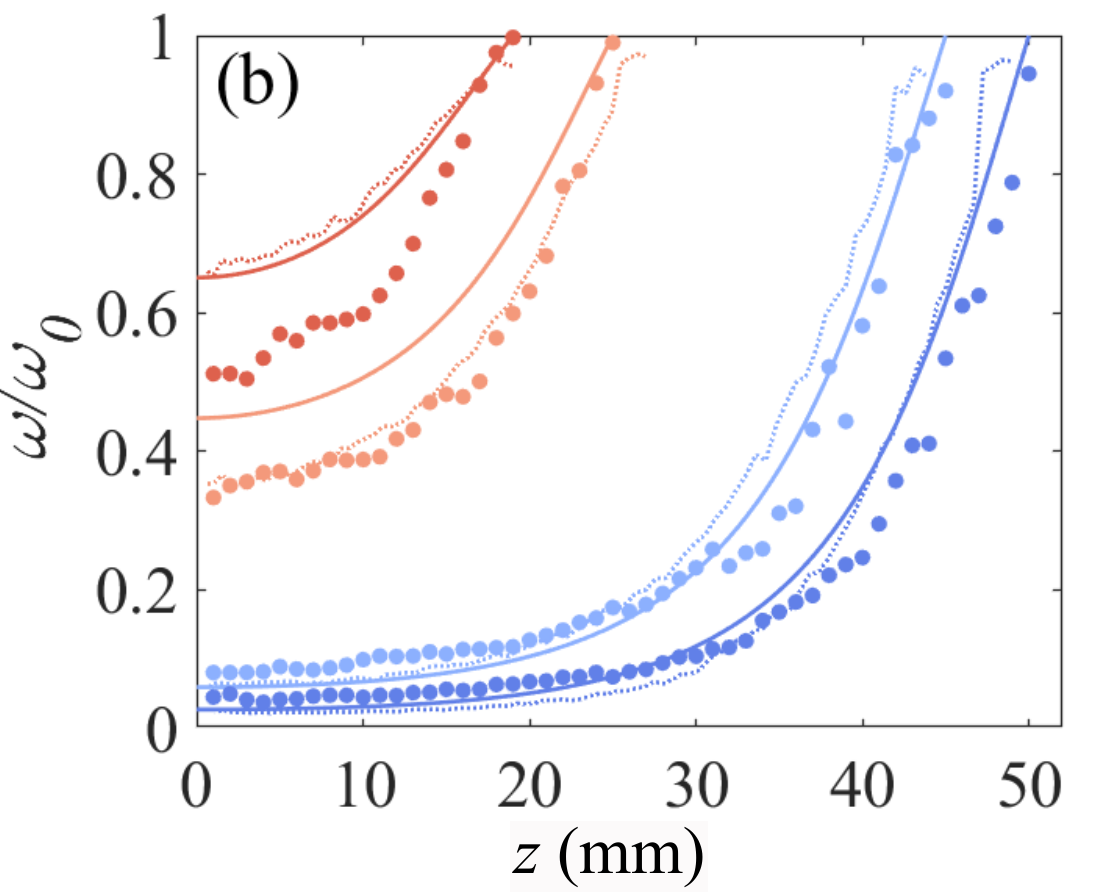}
\caption{Quantitative comparisons between experimental measurements (circles), DEM simulations (dotted lines), and NGF model predictions (solid lines) for flow in the SBSC with $H/R_S$=0.42 \textcolor{Red1}{$\CIRCLE$}, 0.59 \textcolor{Red2}{$\CIRCLE$}, 0.95 \textcolor{Blue2}{$\CIRCLE$}, and 1.05 \textcolor{Blue1}{$\CIRCLE$}. (a) Mid-plane ($z=H/2$) angular velocity comparison as a function of $r/R_S$ for four different $H/R_S$ ratios. (b) Corresponding angular velocity comparisons as a function of $z$ at $r/R_S=0.8$.}
\label{NGF}
\end{figure}
 
To go beyond a qualitative spatial comparison of the flow fields generated by the different methods, Fig.~\ref{NGF} illustrates the quantitative angular velocity profiles extracted from the mid-plane $z=H/2$ (Fig.~\ref{NGF}a) and at a constant radial position $r = 0.8R_S$ (Fig.~\ref{NGF}b) of the 2D profiles of Fig.~\ref{xxxflowprof}. In the experimental data (circles), we observe exceptionally wide shear bands which do not reach $\omega$/$\omega_0=1$ along the mid-plane, even in shallow layers at the lowest height of  $H=15$\,mm. By increasing $H$, the shear band becomes broader and moves inward from a vertical to a horizontal position. From the shallow layer to the deep layer, $\omega$/$\omega_0$ decreases from $\approx 0.8$ to 0.3 at $z=H/2$ and $r\rightarrow 0$; however, the ratio never reaches 0. Figure~\ref{NGF} also illustrates that the predictions of the DEM simulations (dotted lines) and the NGF model (solid lines) are consistent with the experimental data for the four selected fill heights $H/R_S$. DEM simulations can predict the shallow layers well, while the NGF model works well for deeper layers of $H/R_S=0.95,1.05$. However, there is a difference of $\approx 0.15$ in the ratio $\omega/\omega_0$ predicted by the NGF model in shallow layers of $H/R_S=0.42$ and $0.59$ as $r\rightarrow 0$ (Fig.~\ref{NGF}a). For the vertical profiles of $\omega/\omega_0$ at constant radius (Fig.~\ref{NGF}b), DEM matches the experiments well, and, while the NGF model predictions match experiments for deep layers, some discrepancy remains for shallow layers.

\definecolor{Red2d}{rgb}{0.84, 0.19, 0.15}
\definecolor{White2d}{rgb}{0.95, 0.95, 0.95}

\textit{Confining pressure effects ---} We observed exceptionally wide shear zones for unconfined, dense hydrogel suspensions, where the only relevant pressure scale comes from the weight of the particles. This buoyancy-compensated stress field is on the order of 1\,Pa or less~\citep{PhysRevLett.128.238002}. This weak intrinsic pressure scale allows us to further test the pressure sensitivity of the flow profiles. We investigated a confined flow structure where we added a compressive normal stress boundary condition $P$ to the top surface to observe how this affects the shear band structure, using experimental MRI, DEM simulation, and NGF continuum modeling methods.

\begin{figure}[!t]
\centering
\includegraphics[width=0.7\textwidth]{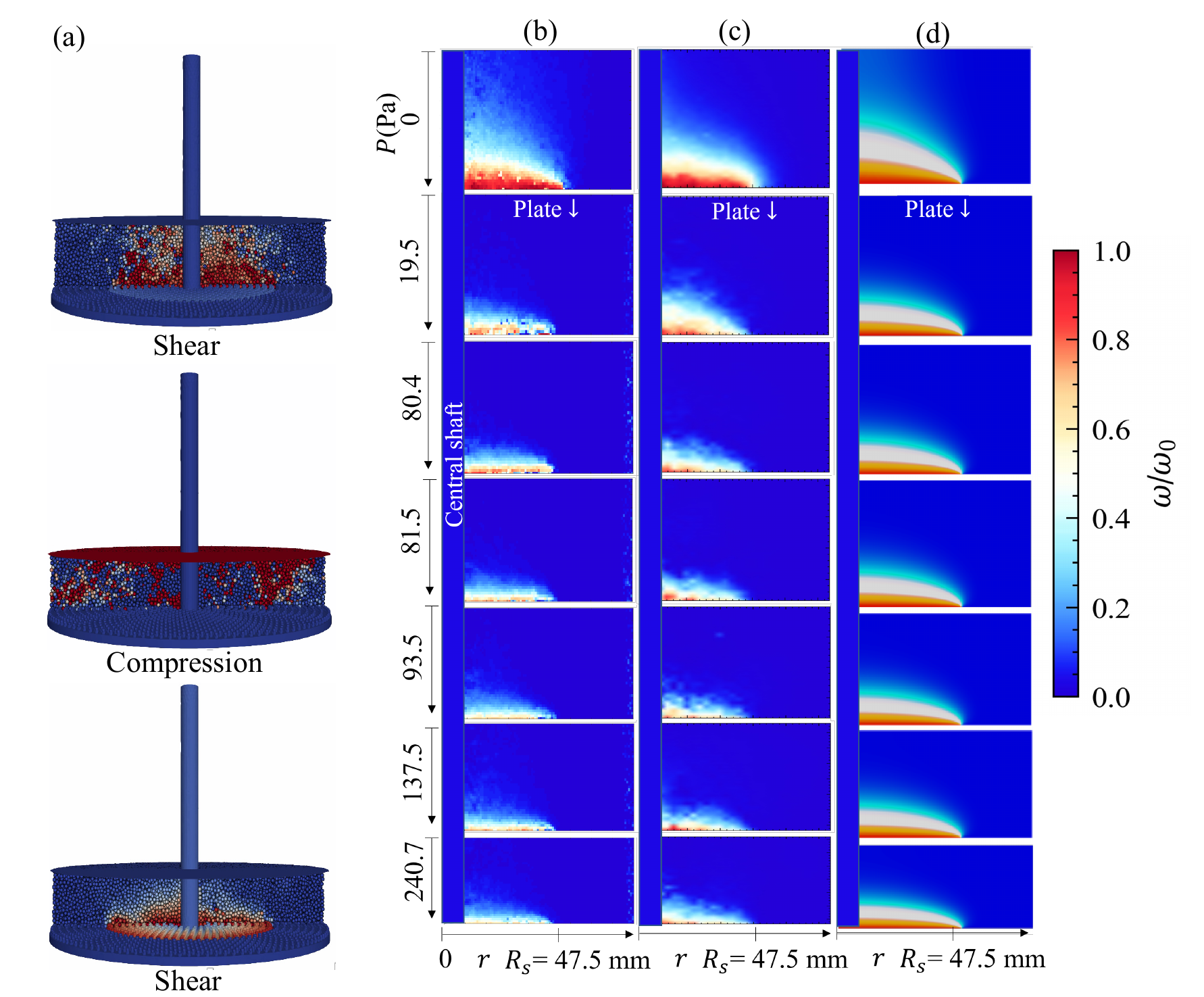}
\caption{(a) Stepwise protocol to measure velocity fields under compression. We shear for a finite strain, then compress, then shear again. Representative images are from DEM results. Qualitative comparisons of non-dimensional, steady-state angular velocity fields from (b) MRI measurements extracted from PIV, (c) DEM simulations, and (d) NGF continuum modeling for dense hydrogel suspensions in a SBSC of fill height $H=50$\,mm with confining pressures of $P=0$, $19.5$, $80.4$, $81.5$, $93.5$, $137.5$, and $240.7$\,Pa. The central shaft is shown as a blue rectangle to the left of the 2D profiles. Static zone (\textcolor{blue}{\SquareSolid}), rotating zone (\textcolor{Red2d}{\SquareSolid}), and shear band (\textcolor{White2d}{\SquareSolid}).}
\label{Stress}
\end{figure}

We performed sets of MRI measurements for filling heights of $H=50, 31,$ and 20\,mm. We focus here on the results from the $H=50$\,mm case, noting that for this case the fixed compression plate least affects the flow structure, as the surface velocity at this $H/R_S$ is already small. The general trends observed are shown in Fig.~\ref{Stress}.  After performing step-wise shear without compressing the packing, we do see the shear band evolve to the surface. However, by performing stepwise shear under compression, the dome becomes much flatter and the flow is more confined to the vicinity of the rotating plate. Fig.~\ref{Stress}a shows the general protocol used to obtain results in both experiment and DEM: we shear the packing, then compress, then shear again immediately. Experimental flow profiles for deep layers of $H=50$~mm from zero to $240.7$\,Pa applied pressure are shown in Fig.~\ref{Stress}b. DEM results in Fig.~\ref{Stress}c with the same top boundary stresses show the same trends as in Fig.~\ref{Stress}b. The results of the non-local continuum model are shown in Fig.~\ref{Stress}d, in which the compressive normal traction $P$ was applied to the top surface, and are likewise consistent with Fig.~\ref{Stress}b. The flow profiles for filling heights of $H=20$ and $31$~mm show a qualitatively consistent behavior and can be found in Appendix~\ref{app:confined}. Our results suggest that adding a small pressure of $\approx20$~Pa is enough to significantly affect the flow profile. By increasing the confining pressure to $\approx240.7$~Pa, the dome becomes increasingly flatter. In the experiments, the dome reaches the point where only one layer of particles moves with the speed of the rotating disk $\omega_0$. This shows the significant effect of the pressure on the width of the shear bands in the SBSC in quasi-static flows, and the decrease in shear-band width with increasing confining pressure is captured by both the DEM simulations, and the NGF continuum modeling. 

\section{Considerations of rate dependence }\label{rate_dependence}

The comparison of SBSC flow profiles for hydrogel suspensions between the experimental MRI, DEM simulation, and NGF modeling approaches relies on the assumption that there are no significant rate dependencies or timescales in the flow. As discussed in Section~\ref{sec:introRD}, the inertial number $I$ and the viscous number $J$ are both much smaller than one, but there are other potential sources of rate dependence that should be excluded. Therefore, we provide evidence, using DEM simulations, that the rate-dependent lubrication forces are not significant in the flow behavior of the hydrogel suspensions at the loading rates applied in the MRI experiments. Then, we probe the rate dependence of both unconfined and confined flows in the SBSC beyond the quasi-static regime using both rheological experiments and DEM simulations. We observe Herschel-Bulkley-type behavior for unconfined suspensions but a deviation from this behavior for the confined suspensions, suggesting that there do exist other relevant timescales, especially for hydrogel suspensions that are under finite compressive stress exceeding the hydrostatic pressure gradient. 

\subsection{Testing the role of hydrodynamic forces in the quasi-static regime}
\begin{figure}[!t]
\centering
\includegraphics[width=0.5\textwidth]{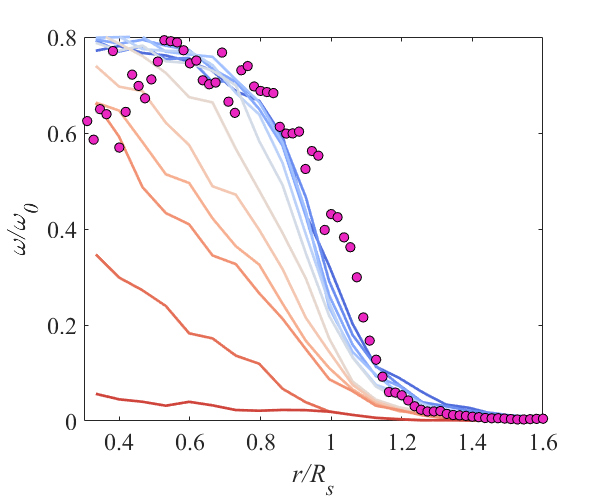}
\caption{Normalized averaged angular velocity profiles along the midplane $(z=H/s)$ for a filling height of $H/R_S=0.32$ from experiment \tikzmycircle[fill=Pink]{3pt} 
and DEM simulations for different ratios of fluid viscosity to water viscosity ($\eta/\eta_{w}$) of 1 \protect\visline, 2 \protect\vislinee, 4 \protect\vislineee, 6 \protect\vislineeee, 8 \protect\vislineeeee, 10 \protect\vislineeeeee, 20 \protect\vislineeeeeee, 30 \protect\vislineeeeeeee, 40 \protect\vislineeeeeeeee, 50 \protect\vislineeeeeeeeee, 100 \protect\vislineeeeeeeeeee, 1000 \protect\vislineeeeeeeeeeee.}
\label{visscale}
\end{figure}

In the hydrogel suspension, the vertical particle pressure gradient is very small as a result of the small density difference between the particles and the surrounding fluid. The small pressure gradient also sets a limit on the total frictional stress on the suspended particles. While we expect that hydrodynamic forces are negligible in the quasi-static regime since the suspending fluid is water with a comparatively low viscosity, it remains to confirm whether the suspended particles experience hydrodynamic forces of comparable or larger magnitude than the frictional contact forces in the slow step-wise shear experiments. To test the role of viscous, rate-dependent contact forces in the flow behavior of the suspended particles, we performed a series of DEM simulations of unconfined flow in a SBSC for a filling height of $H/R_S=0.32$, in which we systematically varied the DEM model fluid viscosity, to determine how such variation would affect the azimuthal flow profiles. The results are shown in Fig.~\ref{visscale} for normalized averaged angular velocity profiles along the midplane $(z=H/2)$ for experimental data (circles) and DEM simulations (solid lines). We observe that in the same conditions under which we performed the MRI experiments, we need to make the model viscosity at least ten times larger to observe a significant shift in the flow profile of the suspension. This confirms that the experimentally realized velocity \textit{during} the application of stepwise shear is in a regime where hydrodynamic forces do not affect the velocity profiles. Note that our DEM model does not capture all hydrodynamic contributions, of which there are several~\cite{Trulsson2017} --- it only accounts for fluid drag via lubrication forces and not the drag from Stokes-type effects that we associate with a finite $J$. In the quasi-static regime, we have assumed that these lubrication forces are the dominant hydrodynamic contribution.

\subsection{Rate dependent flows in unconfined hydrogel suspensions}
\begin{figure}[!t]
\centering
\includegraphics[width=0.5\textwidth]{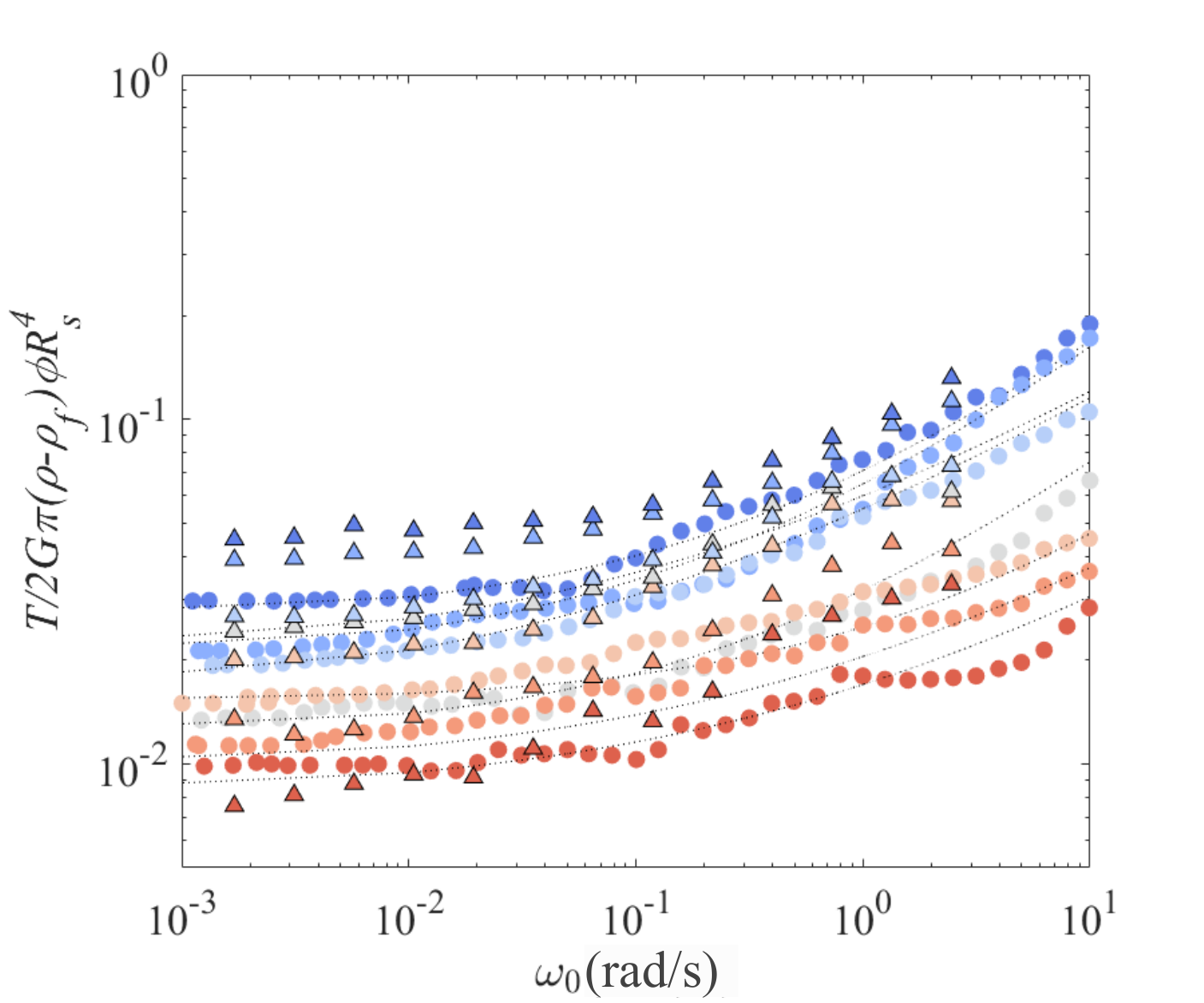}
\caption{The average driving torque $T$ as a function of the shear rate $\omega_0$ for unconfined flows in the SBSC at different filling heights, $H=15$~mm \textcolor{Red1}{\CIRCLE}, 20~mm \textcolor{Red2}{\CIRCLE},  24~mm \textcolor{Red3}{\CIRCLE}, 28~mm \textcolor{White}{\CIRCLE}, 31~mm \textcolor{Blue3}{\CIRCLE}, 38~mm \textcolor{Blue2}{\CIRCLE}, and 46~mm \textcolor{Blue1}{\CIRCLE}, from experiments ($\CIRCLE$) and DEM simulations ($\triangle$). The dotted lines are fits to the HB equation \eqref{HB}. The exponent $n$ is equal to $0.5$ for all filling heights.}
\label{figure8}
\end{figure}

The rate dependence of the flow behavior is an important ingredient in continuum models, as discussed above for the NGF model in, e.g., Eq.~\eqref{local_rheo}; however, extending the NGF model to capture the rate dependence of dense hydrogel particle suspensions is beyond the scope of this work. That said, we provide experimental data that can support future continuum model development. The MRI methods only allow us to probe rate-independent flow profiles, but rate dependence may be probed by rheological measurements. We measured the average driving torque to maintain a certain rotation rate of the disk in the SBSC geometry. Our aim was to probe the relation between the torque $T$ and the rotation rate $\omega_0$ in the confined and unconfined dense hydrogel suspensions. For unconfined cases, based on the pressure-controlled local rheology mentioned in Eq.~\eqref{local_rheo}, we expect an approximately Herschel-Bulkley (HB) type $T(\omega_0)$ of the following form:

\begin{equation}
\label{HB}
    T=T_0+k\omega_0^{n}, 
\end{equation}
where $n\approx 1$ for frictional grains, and $n\approx 0.5$ for frictionless grains (or somewhat smaller as mentioned in Section~\ref{sec:NGF} based on prior works \citep{peyneau2008,srivastava2021}). For confined cases discussed in the next section, based on prior work on constant-volume hydrogel suspension rheology~(\cite{workamp2019contact}, Fig.~5a), in which the ratio of the confining stress and driving stress was observed to be constant in the quasi-static regime, and the limited amount of dilatancy in sheared frictionless suspesions \citep{peyneau2008}, we expect the HB form to also be applicable to the constant-volume, confined suspension experiments. In any case, we focus primarily on the exponent of the HB fit here. We hence tested whether the driving torque upon changing the disk rotation rate can be well fitted to Eq.~\eqref{HB} for all the hydrogel suspension cases tested here. Figure~\ref{figure8} shows the average driving torque $T$ as a function of the driving rate for unconfined flows in the SBSC at different filling heights $H$ for both experiments (circles) and DEM simulations (triangles). We normalize all torque data $T$ with the geometric ingredients for dissipation:
\begin{equation}
T \propto 2G\pi\left(\rho-\rho_f\right)\phi R_S^4. \label{eq:torque}   
\end{equation}
Here, $\phi$ is the packing fraction of the material, which we assume to be 0.64 as is typical for unconfined, disordered frictionless particle packings~\cite{peyneau2008}, although polydispersity can make this value somewhat higher. The trends in the $T$($\omega_0$,$H$) plots are similar for all $H$. First, we observe a rate-independent regime at small $\omega_0$, which corresponds to the range where we performed the MRI measurements discussed in the previous sections. Moreover, the  driving torque increases with $H$. The torque becomes rate dependent for $\omega_0\approx0.01$ -- $0.11$\,s$^{-1}$, and for higher rates, the torque increases with $\omega_0$. Across the whole range of rates, the experimental data can be well fitted to Eq.~\eqref{HB} with the exponent $n=0.5$ (dotted lines). This behavior has been observed for frictionless particles~\cite{PhysRevLett.105.088303, MASON1996439}. We note that the uncertainty of this exponent is large, i.e., a slightly lower exponent would also be consistent with the experimental data, bringing it in the range of the rheological exponent $\alpha \simeq 0.37$, consistent with earlier work~\cite{peyneau2008,srivastava2021}.  



\DeclareRobustCommand\dotted{\tikz[baseline=-0.6ex]\draw[thick,dotted] (0,0)--(0.44,0);}

\subsection{Rate dependent flows in confined hydrogel suspensions}\label{sec:ratedep}

Combining pressure and rate dependence presents additional challenges and cannot be probed via MRI-PIV. We can, however, still characterize the driving torque as a function of both rate and applied pressure. We tested the pressure dependence of the driving torque $T$($\omega_0$,$P$) at a constant filling height of $H=50$\,mm (Fig.~\ref{Torque-P}). We systematically changed the confining pressure at the upper boundary by displacing a porous plate in $2$\,mm steps. We then performed the same shear measurements as before and measured the average driving torque $T$ influenced by $P$, which is plotted using symbols in Fig.~\ref{Torque-P}(a). The rheological behavior is influenced by both the gravitationally induced hydrostatic pressure and the extra applied pressure. We observed a total change in the trend of $T$($\omega_0$) as we increased the confined pressure $P$. At $H=50$~mm before compression, where the only relevant pressure scale is the local hydrostatic pressure, we again see the HB fluid behavior of the suspension with $n=0.5$ (dotted lines in Fig.~\ref{Torque-P}(a)). As we increased the pressure, at $P=80.4$\,Pa, the exponent decreased to $n=0.4$. Surprisingly, a further increase in confining pressure makes $T(\omega_0)$ deviate from HB behavior. At $P=402.7$\,Pa, the torque becomes completely power-law-like with an exponent of $n=0.2$. See Fig.~\ref{Torque-P}(c) for the variation in the fitted HB exponent with $P$ (circles with dotted lines as a guide for the eye). We note that the overall torque level does not grow linearly with $P$. For the regime with power-law flow behavior, as shown in Fig.~\ref{Torque-P}(d), we observe that $T(P)$ measured from experiments at $\omega = 0.01$rad/s (circles) is consistent with $T \propto P^{0.5}$ (dashed line). Note that here the packing density $\phi$ has also changed significantly, as at this pressure the applied strain was about 30\%. See the inset of Fig.~\ref{Torque-P}(c) for the pressure dependence of the volume fraction as extracted from the piston position.. We included the variation of $\phi$ in Eq.~\eqref{eq:torque} for normalizing the obtained torque values.

\definecolor{Blue1}{rgb}{0.38, 0.5, 0.91}
\definecolor{Blue2}{rgb}{0.55, 0.69, 0.996}
\definecolor{Blue3}{rgb}{0.72, 0.81, 0.98}
\definecolor{Blue4}{rgb}{0.9, 0.93, 0.99}
\definecolor{White}{rgb}{0.863, 0.867, 0.867}
\definecolor{Red4}{rgb}{0.97, 0.9, 0.87}
\definecolor{Red3}{rgb}{0.957, 0.77, 0.68}
\definecolor{Red2}{rgb}{0.957, 0.6, 0.482}
\definecolor{Red1}{rgb}{0.87, 0.38, 0.3}
\definecolor{Red0}{rgb}{0.78, 0.34, 0.27}


\begin{figure*}[!t]
\centering
\includegraphics[width=0.7\textwidth]{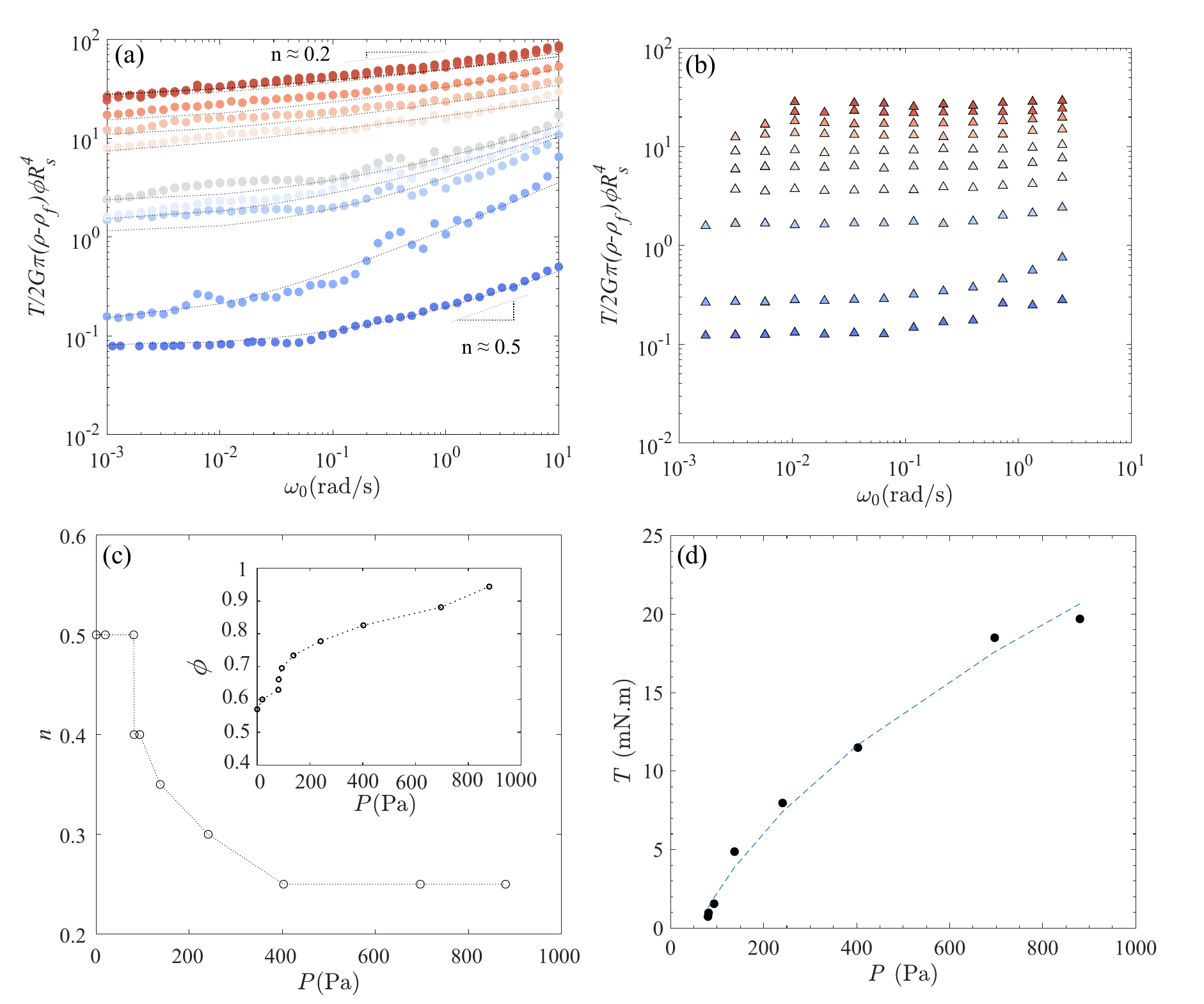}
\caption{The average driving torque $T$ as a function of the driving rate $\omega_0$ at constant filling height of $H=50$\,mm and different confined pressures $P$ of $0$\,Pa \textcolor{Blue1}{\CIRCLE}, $19.5$~Pa \textcolor{Blue2}{\CIRCLE}, $80.4$~Pa \textcolor{Blue3}{\CIRCLE}, $81.5$~Pa \textcolor{Blue4}{\CIRCLE}, $93.5$~Pa \textcolor{White}{\CIRCLE}, $137.5$~Pa \textcolor{Red4}{\CIRCLE}, $240.7$~Pa \textcolor{Red3}{\CIRCLE}, $402.7$~Pa \textcolor{Red2}{\CIRCLE}, $696.7$~Pa \textcolor{Red1}{\CIRCLE}, and $880$~Pa \textcolor{Red0}{\CIRCLE} from (a) experiments ($\CIRCLE$) and (b) DEM simulation ($\triangle$). We fitted the experimental data with the HB model of Eq.~\eqref{HB} ($\dotted$). (c) Change in the HB exponent $n$ versus confining pressure $P$ extracted from experiments (circles with dotted lines as a guide for the eye). Here, the HB  exponent $n$ decreases from $\approx0.5$ in the unconfined flow structure to $\approx0.25$ in the confined flow structure. The inset shows the corresponding volume fraction $\phi$ as a function of $P$. (d) Torque versus confining pressure from experiments at a constant rotation rate of 0.01 rad/s ($\CIRCLE$). The data are fitted by $ T = 9.4\cdot 10^{-4} \sqrt{P} - 7.11 \cdot 10^{-3} $ (dashed line). We emphasize that the manual rotation during MRI scanning was conducted at 0.02 rad/s.}
\label{Torque-P}
\end{figure*}

\section{Discussion}\label{sec:discussion}

\subsection{Shear zone width}

The observed very large shear zone width is an essential flow feature to understand. It is precisely this mesoscopic length scale that seems to depend not only on the particle scale but also the stress field in cases in which ``non-local'' features may be distinguished. In the NGF approach, the difference between the stress ratio and the static yield point sets this length scale through \eqref{eq:cooperativity_length}, so the local pressure and shear stresses are essential components to test for influence. Likewise, in granular temperature models, the diffusivity of the temperature may not be a constant \cite{Berzi2024}.

To understand how local stresses can set the widths of shear zones in dense hydrogel particle suspensions, one should consider several factors. First, since the hydrogel particle--fluid mixture is nearly density-matched, the particle buoyancy nearly balances gravity, and the magnitude of the hydrostatic pressure gradient is small but non-zero. Moreover, for the unconfined cases with zero top pressure, the maximum hydrostatic pressure is also small. Therefore, in unconfined cases, the particle Young's modulus, while small compared to, for example, that of glass beads, is still significantly greater than the maximum pressure, i.e., $\phi(\rho-\rho_f)GH/E_p\ll 1$, and the softness of the hydrogel particles is expected to have a secondary effect on the unconfined flow fields. Second, for unconfined cases in the quasi-static regime, straightforward dimensional analysis suggests that the pressure gradient itself does not affect flow fields because the top pressure is zero, i.e., $P/(\phi(\rho-\rho_f)GH)=0$, and the driving rate is small, i.e., $\omega_0 d/\sqrt{GH} \ll 1$. Third, however, hydrogel particle suspensions have a low particle-particle friction coefficient $\mu_{p-p}$ but still have a finite effective friction coefficient $\mu_s$, due to anisotropy effects~\cite{rothenburg1989analytical}. Moreover, velocity fluctuations in the suspension, known to set the effective ``temperature'' and hence affect the size of the shear zone~\cite{kamrin2019non}, may be influenced by a reduction in particle-particle friction making relative motion of particles at contacts easier, so the low particle-particle friction coefficient in dense hydrogel particle suspensions is expected to be a key microscopic factor affecting the widths of shear zones in unconfined cases. 

That said, the relative effects of (1) the effective friction coefficient $\mu_s$ and (2) the spatial propagation of velocity fluctuations on the large widths of shear zones in hydrogel particle suspensions with low particle-particle friction coefficient is not immediately obvious, but insights may be gained from the NGF model simulations. The effective static yield value of $\mu_s=0.1$ utilized in the NGF model remains much larger than the contact friction coefficient, e.g., $\mu_{p-p}=0.01$ in the DEM simulations, but smaller than $\mu_s$ for frictional grains, e.g., $\mu_s=0.37$ for frictional spheres \cite{zhang2017microscopic}. As a result, the stress ratio $\mu$ is near or just above $\mu_s$ in a large spatial region above the rotating plate in the SBSC, leading to a large value of the cooperatively length $\xi$ in this region and enhanced nonlocal effects. In the NGF model, the spatial propagation of velocity fluctuations is accounted for through the parameter $A$, which, here, was estimated to be $A=0.4$ for hydrogel particles. This value is similar to values previously determined for frictional spheres, e.g., $A=0.43$ in \citet{zhang2017microscopic}, suggesting that the mechanism of propagation of velocity fluctuations is not strongly affected by the reduced particle-particle friction coefficient in dense hydrogel particle suspensions. We note that this is somewhat at odds with the work of \citet{kamrin2014effect}, who found a modest increase in $A$ as the particle-particle friction coefficient decreases in stiff simulated disks. Further work is needed to probe the effect of $\mu_{p-p}$ on the NGF parameters $\mu_s$ and $A$ for spheres, but it is clear that the very large shear zone widths observed in unconfined flows of dense hydrogel particle suspensions in the SBSC may be rationalized through the lens of the NGF model. 

For the confined flows (i.e., $P/(\phi(\rho-\rho_f)GH)\ne 0$), the additional pressure provided by the compressive normal stress on the top surface forces the local ratio of shear to normal stress to be much lower than $\mu_s$ in the whole SBSC except the region just above the rotating plate and hence further away from the divergence of the cooperativity length $\xi$ in the NGF model, reducing the predicted shear band width significantly. While it is encouraging that the flow field characteristics predicted by the NGF model qualitatively match this trend in the MRI measurements at elevated confining pressures, there are several effects neglected in the NGF model that prevent a more quantitative comparison. Most importantly, the current form of the NGF model used in this work neglects the effect of particle softness $P/E_p$, which has been shown to affect the shear zone width in the SBSC \citep{Singh2024} and is expected to become more important in confined flows with enhanced $P$. Prior work on simulated disks \citep{favier2017rheology} explored the effects of particle friction $\mu_{p-p}$ and softness $P/E_p$ on the static yield value in homogeneous shear $\mu_s$ and observed a gentle decrease in $\mu_s$ for softer particles. To enable more quantitative comparisons in confined flows, it remains to elucidate the coupled effects of particle friction $\mu_{p-p}$ and softness $P/E_p$ on both of the NGF model parameters, $\mu_s$ and $A$, and to incorporate these effects into the NGF model, which we leave for future work. Moreover, additional rate-dependent complexities, such as the slow mechanical relaxation of the hydrogel material itself, are not accounted for in the NGF model and could potentially affect the size of shear zones in the quasi-static regime.

\subsection{Rate dependence}

We have established that the rheological behavior of an unconfined hydrogel suspension flow under steady state, constant driving rate conditions is satisfactorily consistent with an HB model. However, when the hydrogel suspension is being compressed, the flow behavior changes profoundly, reaching weak power-law flow curves. These observations are significant for two reasons. First, the DEM models do not reproduce the experimentally-observed power-law flow behavior well, especially at high confining pressures (compare Fig.~\ref{Torque-P}(a) and Fig.~\ref{Torque-P}(b)). This discrepancy suggests that the DEM models are missing an essential piece of physics, e.g., in the contact model. One aspect that thus far has not been included in hydrogel modeling is the time dependence of the hydrogel mechanics: despite the presence of a standard spring-dashpot timescale in the contacts, in a quasi-static flow a classical DEM contact model will not show any rate dependency in the stress response. We speculate that this is because such a contact model only defines a single time scale, namely the contact time, which is shear rate independent. Therefore, the used spring-dashpot contact model is unable to define a second time scale that is, for example, related to the particle material's response to deformation. Consequently, we suggest that it is the slow, viscoelastic relaxation of the hydrogel material that adds an additional timescale (or a second timescale to the contact force), which collectively may reproduce the weak power-law behavior. The similarity of the power-law behavior under confinement observed in the current submersed suspensions bears striking resemblance to the power-law flow behavior observed in adhesive hydrogel packings~\cite{Farmani2025}, suggesting that strong contact deformations perhaps combined with the poro- and viscoelastic nature of hydrogels brings such dynamics about. Second, the power-law flow behavior under constant driving rate conditions is remarkable given the transient creep observations in compressed hydrogel packings exposed to constant stress from an intruder~\cite{Dijksman2024,PhysRevLett.128.238002}. Can these observations be related via, for example, an exponential confinement stress dependence in a prefactor of the power-law flow behavior? Our data is not sufficient to answer these questions, which we leave for future work.

\section{Conclusion}\label{sec:conc}
We have performed experiments as well as discrete element method and continuum-based modeling on the flow behavior of near-density matched, soft frictionless hydrogel suspensions. We have shown that in the rate-independent regime, NGF modeling can still capture both the experimental observations in bulk flow profiles, and the predictions of DEM. These results extend the validity of non-local modeling into materials in which the propagation mechanism for ``fluidization'' is not the same as for nearly rigid particles. We have additionally confirmed that the pressure sensitivity of dense hydrogel particle suspensions is consistent with that predicted in continuum-level NGF modeling. Our work suggests that non-local models generally require stress-dependent closure equations that determine the length scale of these gradient-based models. We have additionally explored the onset of rate dependent flow behavior for hydrogel suspensions, finding broad agreement with DEM results in the case of weak confinement. Rate dependence in strong confinement yields a power-law flow behavior for which we speculate that viscoelastic hydrogel contact mechanics is the culprit.

\begin{acknowledgments}
ZF, NG, SR, JW, RS, JAD acknowledge funding received from the European Union's Horizon 2020 research and innovation programme under the Marie Sk\l{}odowska-Curie grant agreement {\sc CALIPER} No. 812638. Support from NAWI Graz by providing access to its HPC computing resource dcluster.tugraz.at is acknowledged by TU Graz researchers. The authors acknowledge Thomas Gerlach and Oliver Speck for providing the MRI machine time, support, and technical discussions.
\end{acknowledgments}

\appendix

\section{Additional 2D flow profiles - Confined}\label{app:confined}

\begin{figure}[!t]
\centering
\includegraphics[width=0.7\textwidth]{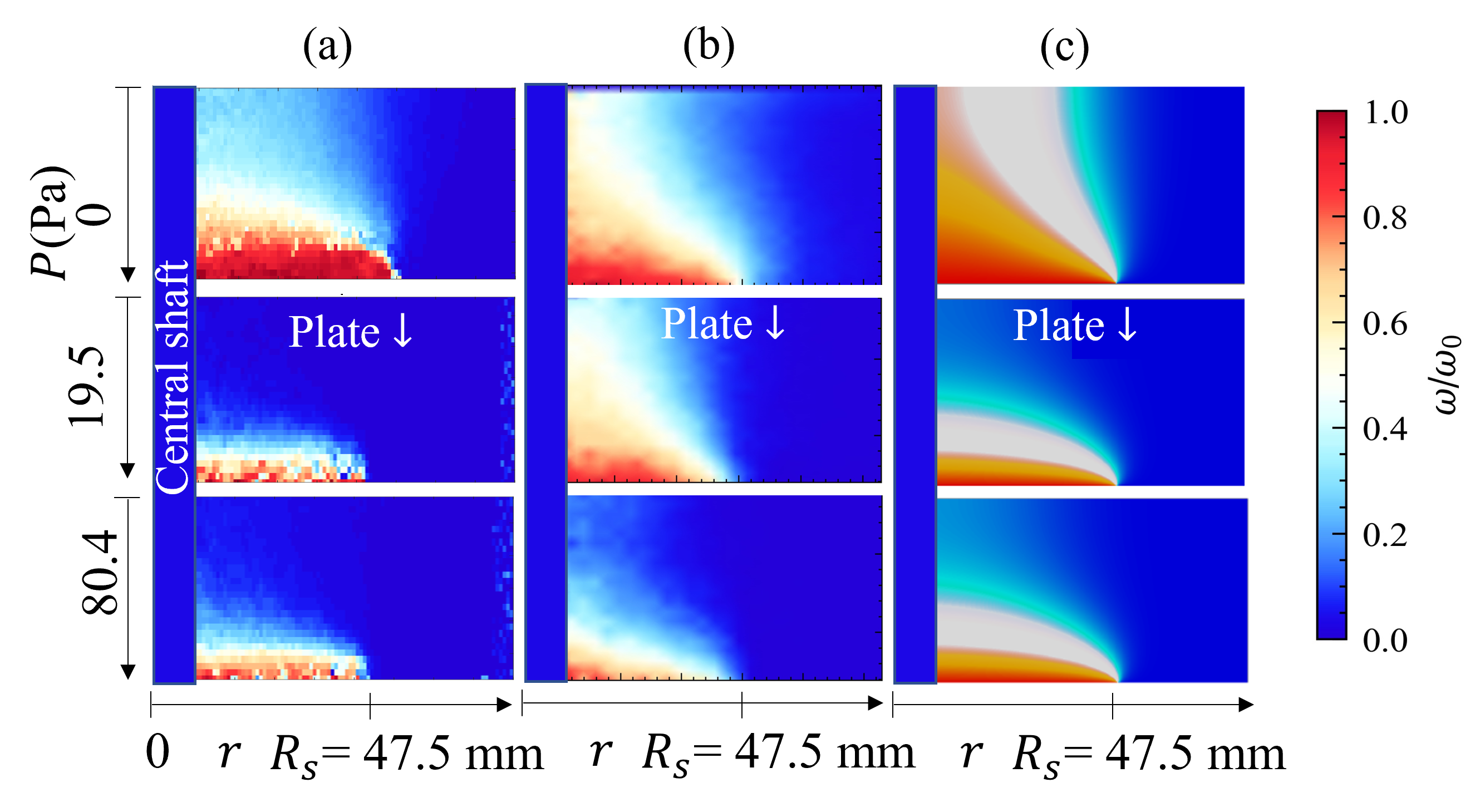}
\caption{Qualitative comparisons of non-dimensional, steady-state angular velocity fields from (a) MRI measurements extracted from PIV, (b) DEM simulations, and (c) NGF continuum modeling for dense hydrogel suspensions in a SBSC of fill height $H=31$\,mm with confining pressures of $P=0$, $19.5$ and $80.4$\,Pa. The central shaft is shown as a blue rectangle to the left of the 2D profiles. Static zone (\textcolor{blue}{\SquareSolid}), rotating zone (\textcolor{Red2d}{\SquareSolid}), and shear band (\textcolor{White2d}{\SquareSolid}).}
\label{figure_comp_31}
\end{figure}

\begin{figure}[!t]
\centering
\includegraphics[width=0.7\textwidth]{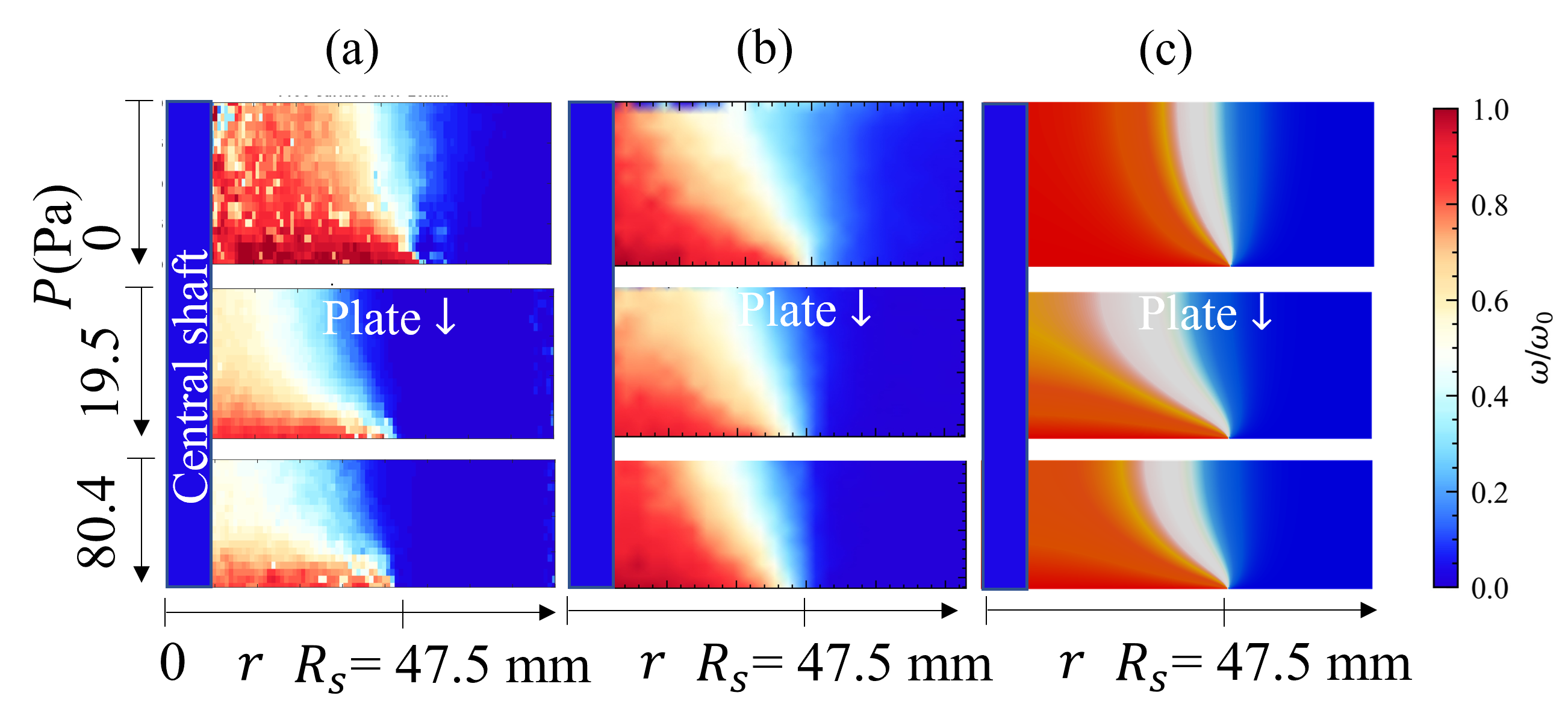}
\caption{Qualitative comparisons of non-dimensional, steady-state angular velocity fields from (a) MRI measurements extracted from PIV, (b) DEM simulations, and (c) NGF continuum modeling for dense hydrogel suspensions in a SBSC of fill height $H=20$\,mm with confining pressures of $P=0$, $19.5$ and $80.4$\,Pa. The central shaft is shown as a blue rectangle to the left of the 2D profiles. Static zone (\textcolor{blue}{\SquareSolid}), rotating zone (\textcolor{Red2d}{\SquareSolid}), and shear band (\textcolor{White2d}{\SquareSolid}).}
\label{figure_comp_20}
\end{figure}

In addition to the flow profiles observed at finite compression for a filling height of $H=50$\,mm (Fig.~\ref{Stress}), we provide for completeness the flow profile information for experiments performed at additional filling heights. In this appendix, we include steady-state angular velocity fields from experiments, DEM simulations, and NGF continuum modeling for fill heights of $H=31$\,mm (Fig.~\ref{figure_comp_31}) and $H=20$\,mm (Fig.~\ref{figure_comp_20}). We chose to focus on the larger fill height in the main body of the article because the top compression plate imposes both a normal stress and an unknown shear traction on the particles in the top layer. In the NGF model predictions, the shear traction applied by the top plate is idealized to be zero. In the case of the largest filling height, the angular velocity in the top layer is minimal, so the shear boundary condition at the top surface has a minimal effect on the flow fields. That said, we can see in Figs.~\ref{figure_comp_31} and \ref{figure_comp_20} that there is still qualitative agreement between the flow fields from experiments, DEM simulations, and NGF modeling in shallower layers and increasing top-plate pressure. 

\DeclareRobustCommand\dotted{\tikz[baseline=-0.6ex]\draw[thick,dotted] (0,0)--(0.44,0);}


\begin{figure}[!t]
\centering
\includegraphics[width=0.5\textwidth]{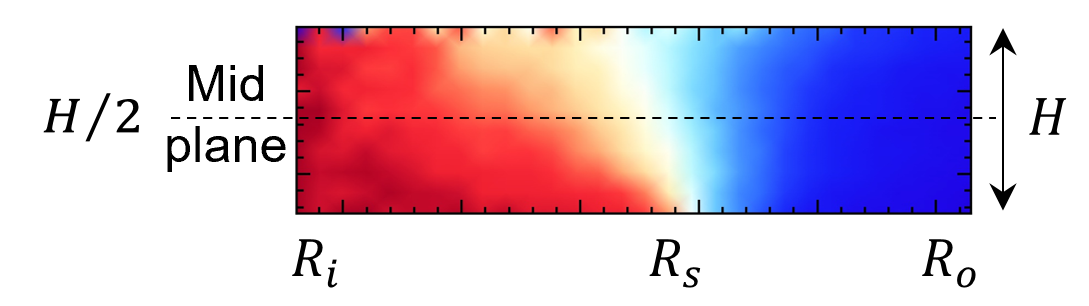}
\includegraphics[width=0.5\textwidth]{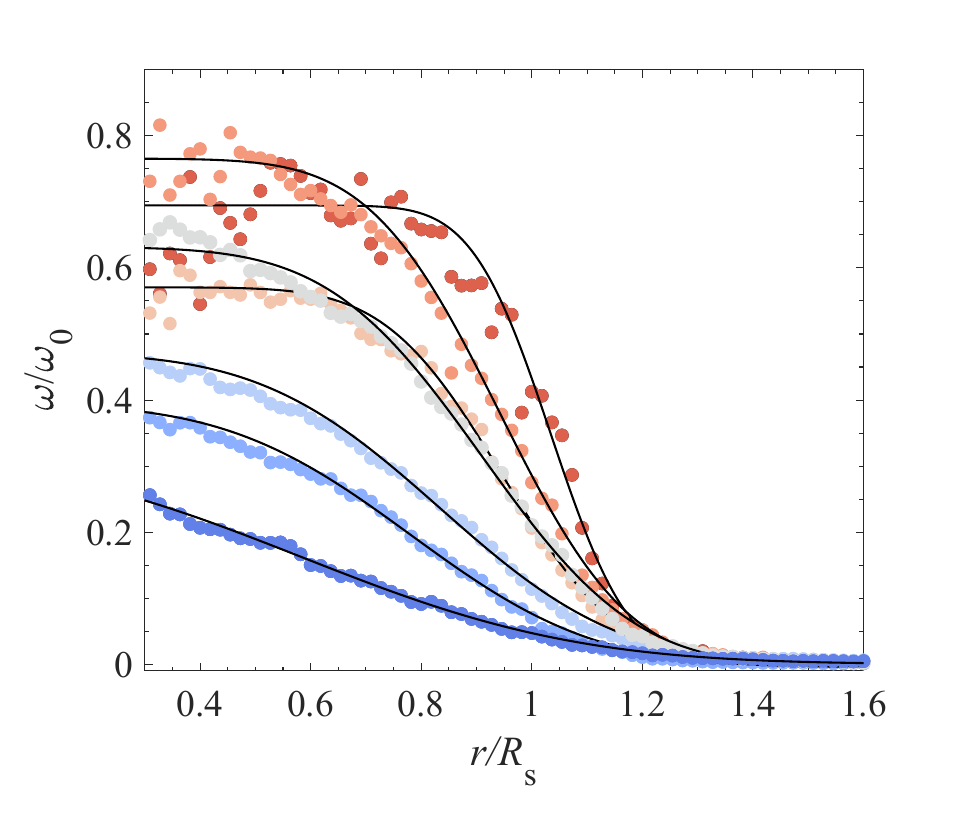}
\caption{Normalized averaged angular velocity profiles (MRI measurements) for a range of $H/R_S=0.31$ \textcolor{Red1}{$\CIRCLE$}, 0.42 \textcolor{Red2}{$\CIRCLE$}, 0.50 \textcolor{Red3}{$\CIRCLE$}, 0.59 \textcolor{White2d}{$\CIRCLE$}, 0.65 \textcolor{Blue3}{$\CIRCLE$}, 0.95 \textcolor{Blue2}{$\CIRCLE$}, and 1.05 \textcolor{Blue1}{$\CIRCLE$}. The plot shows the normalized averaged angular velocity $\omega$/$\omega_0$ as a function of radius $r$ along the mid-plane $z = H/2$. For all cases, $R_S$ is 47.5\,mm. All the profiles are fitted with an error function, Eq.~\eqref{ERFC}, shown by \protect\blackline. Note that the center region has larger uncertainties in the estimate of the angular velocity due to PIV limitations.
}
\label{figure6}
\end{figure}

\section{Shear band characterization in unconfined flows}\label{app:shearband} 
\begin{figure}[!t]
\centering
\includegraphics[width=0.9\textwidth]{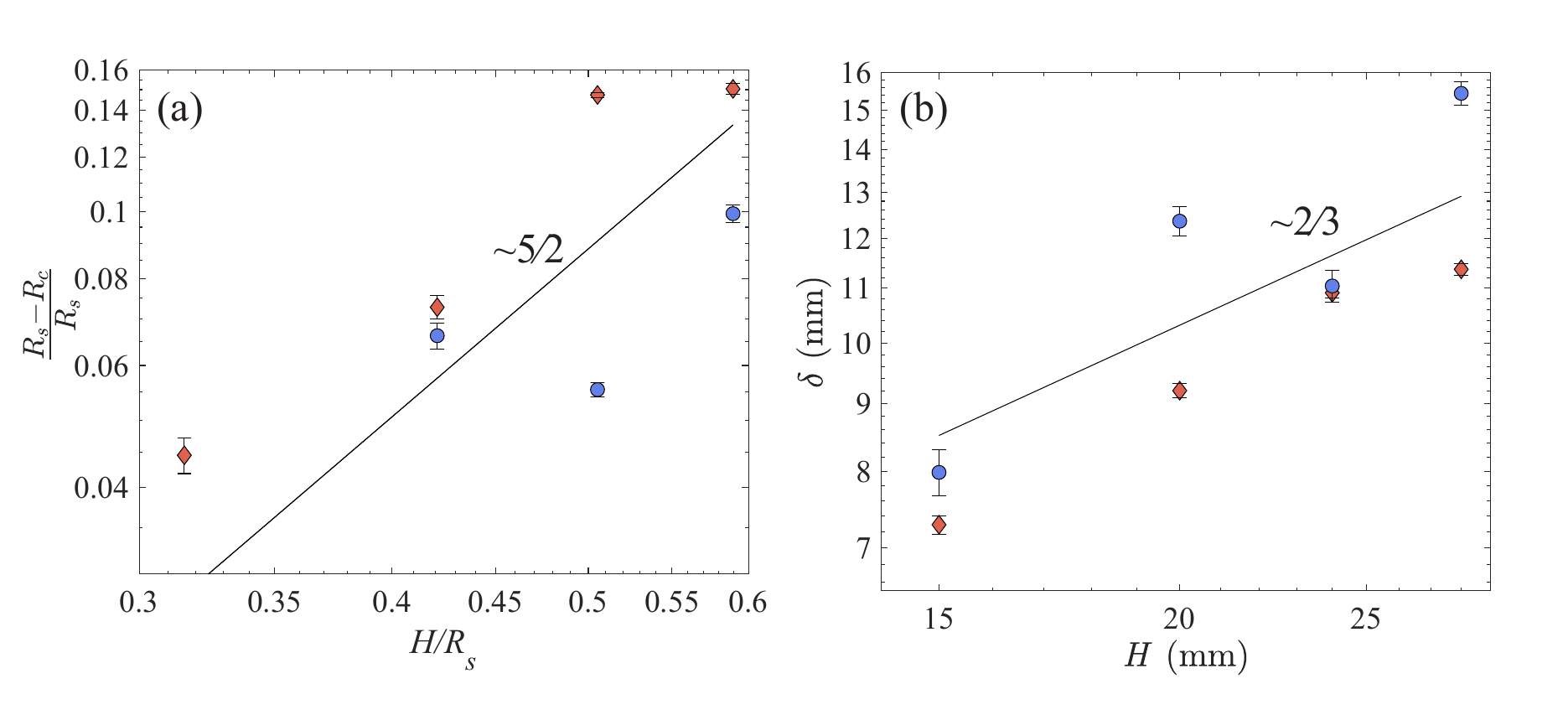}
\protect\caption{Shear band characterization in unconfined flows of dense hydrogel suspensions extracted from fitting Eq.~\eqref{ERFC} to the velocity profiles of Fig.~\ref{figure6} from experiments \protect\tikzmycircle[fill=Blue1]{3pt} and DEM simulations \protect\textcolor{Red1}{\protect\mydiamond}. (a) Dimensionless displacement of the shear band center $({R_S-R_c})/{R_S}$ as a function of the ratio of filling height to split radius $H/R_S$. For reference, the literature standard~\cite{dijksman2010granular} fit $({R_S-R_c})/{R_S}=0.5(H/R_S)^{5/2}$ is shown as a line (\protect\blackline) (b) Shear band width $\delta$ as a function of filling height $H$. The fit $\delta=1.4(H)^{2/3}$  is shown as a line (\protect\blackline), illustrating the literature scaling \cite{fenistein2004universal}. Plots are in logarithmic scale. One experimental value in (a) is negative and does not appear on the graph. Error bars represent fit errors of individual runs, not an estimate of all systematic and statistical sources of errors.}
\label{figure14}
\end{figure}

As shown in Fig.~\ref{figure6}, we characterize shear bands in unconfined flows through their centers $R_c$ and widths $\delta$ by fitting an error function (solid lines) to the experimental velocity profiles (circles) along the mid-plane ($z=H/2$). All the velocity profiles can be fitted satisfactorily by an error function \cite{fenistein2004universal}: 
\begin{equation}
\label{ERFC}
\omega\left(r\right)=\frac{A_0\omega_0}{2}\left[{\rm erf}\left(\frac{R_c-r}{\delta}\right)+1\right],
\end{equation}
with fitting parameters $\{R_c,\delta,A_0\}$. By fitting the velocity profiles of Fig.~\ref{figure6} with Eq.~\eqref{ERFC}, we extract $R_c$ and $\delta$ and plot the change in $R_c$ and $\delta$ with respect to the filling height $H$ in Fig.~\ref{figure14} for experiments (circles) and DEM simulations (diamonds). Based on previous studies~\cite{fenistein2003wide}, the only relevant length scales for $R_c$ are the geometric scales $H$ and $R_S$, and it has been shown that the dimensionless displacement of the shear band center, $1-R_c/R_S$, should be a function of $H/R_S$. The variation of $R_c$ with $H$ in our system is consistent with a simple scaling law obtained previously in the literature \cite{fenistein2003wide}: $1-R_c/R_S\sim (H/R_S)^{5/2}$ (solid line in Fig.~\ref{figure14}(a)).

Data on the width of the shear band $\delta$ (Fig.~\ref{figure14}(b)) indicate that the shear band width grows non-linearly with the filling height $H$ and is consistent with the power-law scaling obtained previously in the literature \cite{fenistein2004universal}: $\delta \sim H^{2/3}$ (solid line in Fig.~\ref{figure14}(b)).
%


\section{Validating Torque Calculations}\label{torque}
\begin{figure}[htbp]
\centering
\includegraphics[width=0.6\textwidth]{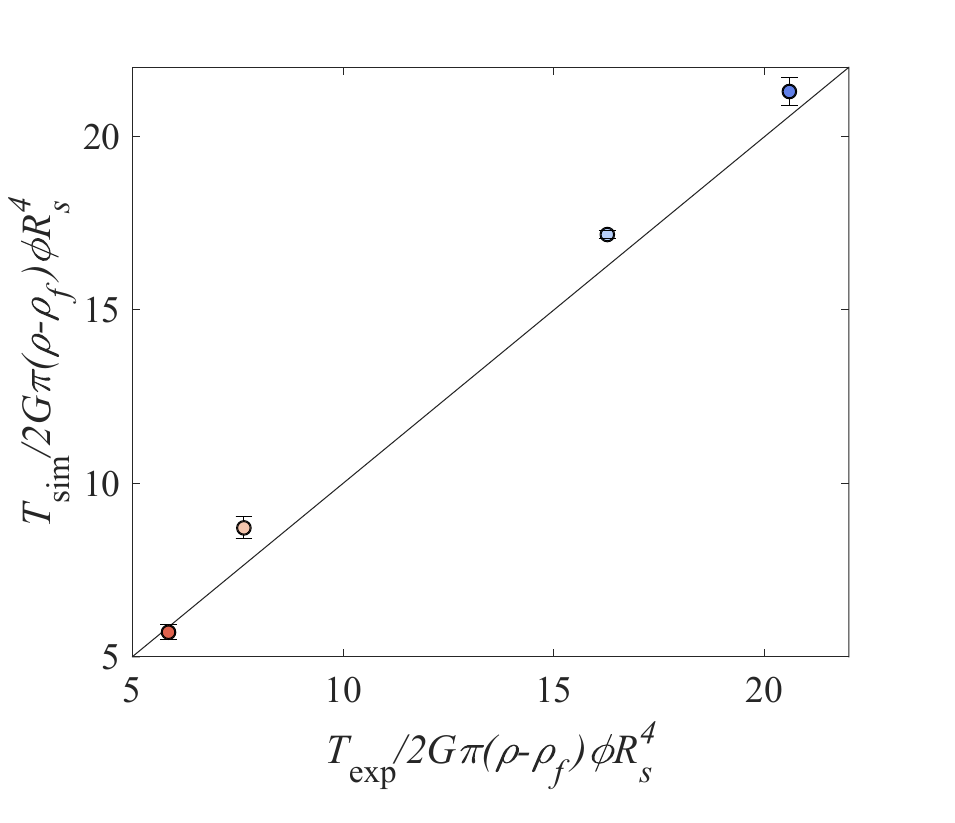}
\caption{ Comparison of the average torques measured in the experiments and calculated in DEM simulations for glass bead particles at four different filling heights of $H=8$~mm \protect\tikzmycircle[fill=Red1]{3pt}, 10~mm \protect\tikzmycircle[fill=Red2]{3pt}, 15~mm \protect\tikzmycircle[fill=Blue2]{3pt}, and 18~mm \protect\tikzmycircle[fill=Blue1]{3pt}. The solid line is the ideal result $T_{\rm sim} = T_{\rm exp}$. Note that $\rho_f$ is the density of the air around the glass beads, which is negligible.}
\label{figure20}
\end{figure}

In order to verify the torque calculation in the DEM simulations, the split-bottom shear test was performed with glass bead particles with a 50:50 (mass-based) binary distribution with a diameter ratio of 1:1.05, where the diameter of smaller particles is 1.9~mm. The DEM properties are summarized in Table \ref{tab:table2}. The test was carried out in the rate-independent regime for 20 values of the angular velocity of the rotating base $\omega_0$ over the range 0.0039--3.9\,rad/s. The setup was filled with particles, settled, and then sheared for 7 rotating steps, as described above. The first two rotation steps were excluded from averaging. The comparison of the DEM simulations and the experimental results is shown in Fig.~\ref{figure20} (circles) and compared with the ideal result $T_{sim}=T_{exp}$ (solid line).

\begin{table}
    \centering
    \caption{DEM model parameters for glass bead particles. Note that the Young's modulus of the glass beads was reported to be in the range of 78 to 85 GPa by the manufacturer, but after sensitivity analysis, it was reduced for the benefit of computational costs.}
    \begin{tabular}{|l|l|}
    \hline
       Parameter  & Value \\
       \hline
         Particle density ($\rho$) & 2500 {kg/m$^3$}\\
         Particle-Particle sliding friction ($\mu_{s,p-p}$) & 0.34\\
         Particle-Wall sliding friction ($\mu_{s,p-w}$) & 0.3\\
         Particle-Particle rolling friction ($\mu_{r,p-p}$) & 0.1\\
         Particle-wall rolling friction ($\mu_{r,p-w}$) & 0.1\\
         Particles' Young's Modulus($E_{p}$) & 7.8 MPa\\
         Wall Young's Modulus ($E_w$) & 2.5 GPa\\
         Coefficient of restitution ($e_{p-p}= e_{p-w}$) & 0.9\\
         Particles' Poisson's ratio ($\nu_p$) & 0.245\\
         Wall Poisson's ratio ($\nu_w$) & 0.3\\
         Gravity ($G$) & 9.81 {m/s$^{2}$}\\
         \hline
    \end{tabular}
    \label{tab:table2}
\end{table}

\section{Obtaining the steady-state form of the NGF model}\label{app:NGFss} Here, we discuss how the non-local rheology, \eqref{nonlocal_seg}, \eqref{eq:local_fluidity}, and \eqref{eq:cooperativity_length}, is obtained from the steady-state approximation of a dynamical PDE. As the starting point, we adopt the following dynamical, Ginzburg-Landau-like PDE governing the granular fluidity, which is slightly modified from the form described in \citet{henann2014continuum} to account for the power-law local rheology:
\begin{equation}\label{eq:dynamic_pde}
t_0 \dot{g}  = A^2d^2 \frac{\partial^2 g}{\partial x_k \partial x_k } - (\mu_s - \mu)g - b \left( \sqrt{\frac{d^2 \rho}{P} }\,\mu g \right)^{\alpha} g,
\end{equation}
where $t_0$ represents the time scale associated with the transient evolution of the fluidity field, and $\{\mu_s, b, \alpha, A\}$  is the parameter set associated with the non-local model --- the parameter $t_0$ does not enter the steady-state form. In the absence of spatial gradients, the steady-state solution of \eqref{eq:dynamic_pde} is denoted as the local solution $g_{\rm loc}$. When $\mu < \mu_s$, there is only one non-negative solution, $g_{\rm loc} =0$, and it is stable. For $\mu > \mu_s$, there are two types of solutions: (1) $g_{\rm loc} = 0$, which is unstable, and (2) $g_{\rm loc} = \sqrt{\frac{P}{{d}^2\rho}} \,\frac{1}{\mu}  \left(\frac{\mu-\mu_s}{b}\right)^{1/\alpha}$, which is the stable solution. The stable solutions give the local fluidity function \eqref{eq:local_fluidity}. 

Then, in the presence of spatial gradients, we limit our analysis of stable solutions to small deviations of $g$ from $g_{\rm loc}$, so that $g=g_{\rm loc} + \delta$, and linearize the PDE \eqref{eq:dynamic_pde} in $\delta$ for $\dot{g}=0$. The two cases are as follows:
\begin{enumerate}
\item[{\bf Case 1}:] For $\mu < \mu_s$, $g_{\rm loc} = 0$, and $g=\delta$. The steady-state form of \eqref{eq:dynamic_pde} then becomes
\begin{equation}\label{eq:dynamic_pde_nonlin}
 A^2d^2 \dfrac{\partial^2\delta}{\partial x_k\partial x_k} = (\mu_s - \mu) \delta + b \left(\sqrt{\frac{d^2 \rho}{P}} \,\mu \right)^{\alpha} \delta ^{1 + \alpha}.
\end{equation}
Neglecting the higher-order term in $\delta$, we obtain the following linearized form  $$A^2d^2 \dfrac{\partial^2 g}{\partial x_k\partial x_k} = (\mu_s - \mu)g.$$
\item[{\bf Case 2}]:] For  $\mu > \mu_s$,  $g_{\rm loc} \neq 0$, and $g = g_{\rm loc} + \delta$. Again, substituting in \eqref{eq:dynamic_pde}, followed by a (straightforward but somewhat lengthy) linearization, we obtain
\begin{equation}\label{eq:dynamic_pde_steady1}
 A^2d^2 \dfrac{\partial^2 g}{\partial x_k\partial x_k} = \alpha (\mu - \mu_s)(g - g_{\rm loc}).
\end{equation}
\end{enumerate}
Putting the two cases together, the steady state form of \eqref{eq:dynamic_pde} becomes
\begin{equation}\label{eq:dynamic_pde_steady2}
 \xi^2(\mu) \dfrac{\partial^2 g}{\partial x_k\partial x_k} = g - g_{\rm loc}(\mu,P).
\end{equation}
where $g_{\rm loc}(\mu,P)$ is given by \eqref{eq:local_fluidity}, and $\xi(\mu)$ is given by \eqref{eq:cooperativity_length}. We note that a similar form for the cooperativity length was obtained by \citet{bocquet2009kinetic} for $\alpha=1/2$ in the case of emulsions.

\clearpage
\bibliography{apssamp}
\bibliographystyle{plainnat}
\end{document}